\documentclass[english,12pt,aps,prd,a4paper,preprintnumbers,floatfix,nofootinbib,showpacs,superscriptaddress, notitlepage]{revtex4-1} 
\usepackage[usenames,dvipsnames]{color}  
\usepackage{graphicx}
\usepackage{setspace}
\usepackage{caption}
\usepackage{amsmath}
\usepackage{amssymb}
\usepackage[normalem]{ulem}
\usepackage{xcolor}
\usepackage{placeins} 
\usepackage{mathrsfs}
\usepackage{verbatim} 
\usepackage{cancel} 
\usepackage{float} 
\RequirePackage[colorlinks=true,urlcolor=BrickRed,anchorcolor=BrickRed,citecolor=BrickRed,filecolor=BrickRed,linkcolor=BrickRed,menucolor=BrickRed,pagecolor=BrickRed,linktocpage=true,pdfproducer=medialab,pdfa=true]{hyperref}

\usepackage{scalerel}
\usepackage{tikz-feynman}
\tikzfeynmanset{compat=1.1.0}
\usepackage{feynmp}
\usepackage{tikzsymbols}
\usepackage{subfigure}

\makeatletter
\def\p@subsection{}
\makeatother

\definecolor{darkred}{rgb}{0.6,0,0}

\definecolor{linkcolor}{rgb}{0,0,0.5}

\def\gsim{\raise0.3ex\hbox{$\;>$\kern-0.75em\raise-1.1ex\hbox{$\sim\;$}}}
\def\lsim{\raise0.3ex\hbox{$\;<$\kern-0.75em\raise-1.1ex\hbox{$\sim\;$}}}

\def\beqn#1{\begin{equation}\label{#1}}
\def\eeqn{\end{equation}}

\def\beqa#1{\begin{eqnarray}\label{#1}}
\def\eeqa{\end{eqnarray}}

\definecolor{denim}{rgb}{0.08, 0.38, 0.74}

\newcommand {\ignore}[1]{}

\def\Z4{$Z_4$ }
\def\O5{$\mathcal{O}_5$ }

\def\321{$\mathrm{SU(3) \otimes SU(2) \otimes U(1)}$ }

 \newcommand{\AddrIISERB}{Department of Physics,
 Indian Institute of Science Education and Research - Bhopal \\
 Bhopal Bypass Road, Bhauri, Bhopal, India}

\begin{document}

\title{\color{BrickRed} 
Dark Matter Escaping Direct Detection Runs into \\
Higgs Mass Hierarchy Problem } 

\author{Praveen Bharadwaj}\email{praveen20@iiserb.ac.in}
\affiliation{\AddrIISERB}
\author{Ranjeet Kumar}\email{ranjeet20@iiserb.ac.in}
\affiliation{\AddrIISERB}
\author{Hemant Kumar Prajapati}\email{hemant19@iiserb.ac.in}
\affiliation{\AddrIISERB}
\author{Rahul Srivastava}\email{rahul@iiserb.ac.in}
\affiliation{\AddrIISERB}
\author{Sushant Yadav}\email{sushant20@iiserb.ac.in}
\affiliation{\AddrIISERB}

\begin{abstract}
  \vspace{1cm}

The current generation of Dark Matter Direct Detection Experiments has ruled out a large region of parameter space for dark matter, particularly in the ($10 - 1000$) GeV mass range. However, due to very low event rates, searching for dark matter in the heavy mass range, $\mathcal{O}$(TeV), is a daunting task requiring even larger volume detectors and long exposure times.
We show that for a broad class of dark matter models of the type that these experiments are searching,  including some of the most popular candidates, the heavy dark matter mass range can be ruled out in its entirety once we take into account the large corrections to Higgs mass imparted by such heavy dark matter as shown below for the singlet scalar dark matter case.
\begin{figure}[h!]
\centering
\includegraphics[width=.6\textwidth] {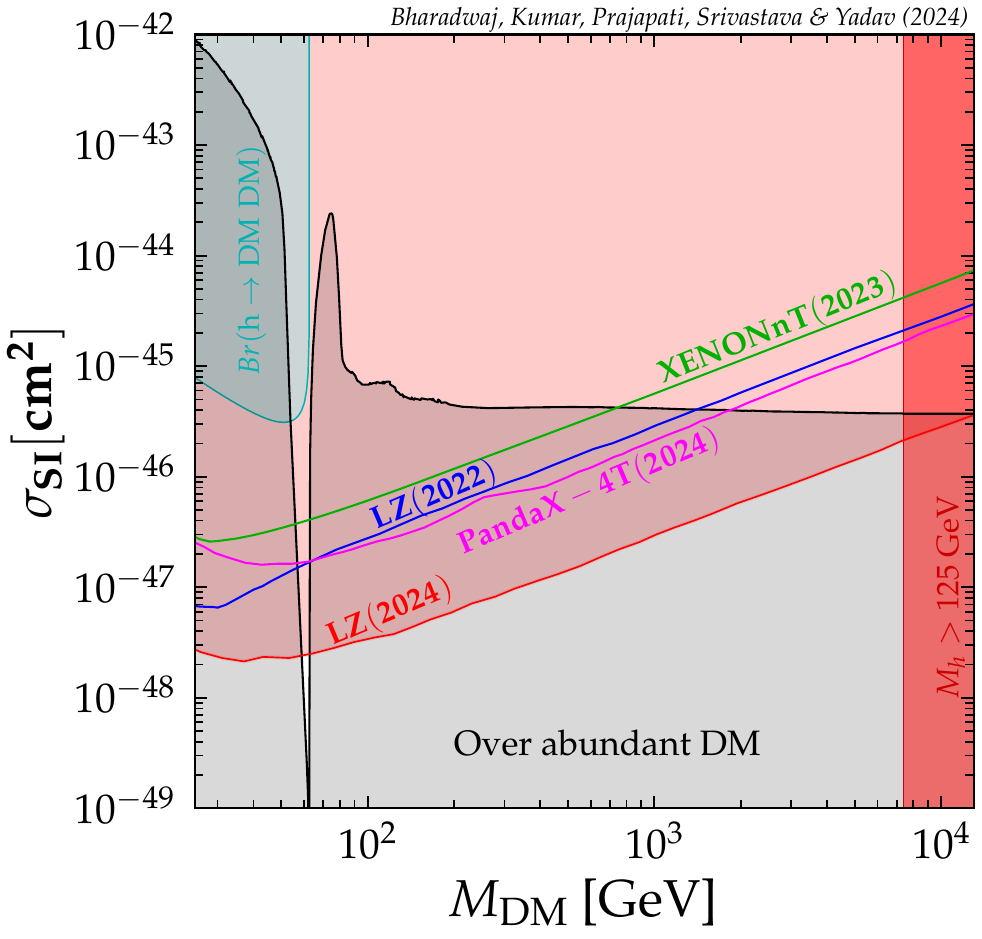}
\label{fig:abs}
\end{figure}
We show that such a limit is applicable to all types of dark matter i.e. scalar, vector, and fermionic, provided they couple directly with Higgs. By taking some simple and well studied dark matter models we show that the latest LZ limits can completely rule out such a dark matter except in a narrow range around $M_h/2$ mass, similar to the example shown in the figure above. 

\end{abstract}
\maketitle



\section{Introduction}
\label{sec:intro}

Observations from cosmology and astrophysics consistently suggest the presence of a non-baryonic form of matter~\cite{Zwicky:1933gu,Rubin:1970zza,Rubin:1980zd,Planck:2018vyg}, commonly referred to as dark matter (DM). The longstanding challenge of understanding the nature and existence of DM provides a compelling justification for the investigation of physics beyond the Standard Model (SM). It is often hypothesized that DM is comprised of an entirely new, cold (non-relativistic), electrically neutral massive particle which interacts very weakly with the SM particles. Numerous extensions to the SM are proposed to account for DM. The most popular being models with weakly interacting massive particles (WIMPs). 

The defining feature of WIMP dark matter is that it is created thermally and 
is in thermal equilibrium with SM particles in the early universe. As the universe expands and cools, the temperature drops, hence the interaction between WIMPs and SM particles decreases. This leads to the thermal ``freeze-out" scenario \cite{Scherrer:1985zt} where WIMPs could no longer annihilate efficiently, leaving behind a stable relic density, consistent with observations~\cite{Planck:2018vyg}. 
Thus, one can obtain a mass range for the WIMPs ranging from few KeV to roughly 100 TeV. The lower mass limit is obtained from the requirement that DM should be cold \cite{Colombi:1995ze,Viel:2013fqw}, while the upper limit comes from Unitarity bounds \cite{Griest:1989wd,Smirnov:2019ngs}.   

The simplicity of these WIMP models, combined with their detectable freeze-out signatures and ongoing experimental searches in wide mass range, has led to a very large chunk of theoretical effort to build WIMP like DM models. Some of the most popular and simple DM models are the various versions of singlet scalar \cite{Cline:2013gha,Profumo:2014opa,Mandal:2023oyh}, inert doublet \cite{Deshpande:1977rw}, scotogenic \cite{Tao:1996vb,Ma:2006km,Rojas:2018wym,Leite:2019grf,Barreiros:2020gxu,Mandal:2021yph,Batra:2022pej,CentellesChulia:2022vpz,Kumar:2023moh,Kumar:2024zfb,CentellesChulia:2024iom}, lepton quarticity \cite{CentellesChulia:2016rms,CentellesChulia:2016fxr,CentellesChulia:2017koy,Srivastava:2017sno}, singlet fermion \cite{Prajapati:2024wuu} etc. 
Most of these models have DM candidates which are color and electromagnetic singlets. They can be charged under weak interaction and typically also couple with Higgs. In particular, if the DM is a scalar particle, then its interaction with Higgs is unavoidable. For these models, their coupling with W, Z, and Higgs plays a crucial role in determining the correct relic abundance through freeze-out and in their direct detection. 

The relic density set by the freeze-out mechanism is directly linked to the strength of WIMP interactions, which are weak but non-zero, making WIMPs detectable via their scattering interactions with detector nuclei in direct detection experiments~\cite{Goodman:1984dc}. Currently,  dark matter direct detection experiments are searching for such WIMPs in the range of few MeVs till TeVs. Searching for WIMPs over such a wide range is a daunting task, particularly for very low mass $M_{\rm{DM}} \lsim \mathcal{O}$ (GeV) and high mass  $M_{\rm{DM}} \gsim \mathcal{O}$ (TeV) ranges. The average local DM density is roughly given by $0.3\, \rm{GeV}/\rm{cm}^3$, so in the low mass range, the DM flux is large. However, the problem is that the small nuclear/electronic recoil imparted by the DM makes the detection challenging and requires special detectors and detection techniques. In the high mass range, we have the opposite problem, i.e. the DM flux becomes very small, leading to low event rates. Ruling out WIMPs in the $(10 - 100)$ TeV mass range will require future detectors with much larger volume and long exposure times~\cite{PANDA-X:2024dlo,XLZD:2024nsu}.

In this work, we systematically analyze the loop corrections to the Higgs mass induced by DM in WIMP dark matter models. We argue that, for a large class of such models, the high dark matter mass range can be ruled out when these corrections are taken into account. As we discuss in Sec.~\ref{sec:higgs-dm-int} and elaborate further in Sec.~\ref{sec:relic}, in large class of DM models, the Higgs-DM coupling is crucial to obtain the correct DM relic abundance. Thus, for a given DM mass, the DM relic abundance fixes the strength of the Higgs-DM coupling. In particular, the larger the DM mass, the larger should be Higgs-DM coupling in order to satisfy the DM relic abundance.
Now, as we discuss in Sec.~\ref{sec:eft_loop} and Sec.~\ref{sec:relic}, the DM-induced loop corrections to Higgs mass are also proportional to the Higgs-DM coupling and DM mass. Therefore, as the DM mass increases, the magnitude of DM-induced loop corrections also increases.
We show that, indeed, for non-SUSY DM models, these corrections can become quite large when the DM mass reaches a few TeV. 
Consequently, this sets an upper limit on DM mass, beyond which Higgs mass can no longer be maintained at its observed value of $125.20 \pm 0.11 $ GeV \cite{ParticleDataGroup:2024cfk} as discussed in Sec.~\ref{sec:eft_loop} and Sec.~\ref{sec:relic} for various types of DM models. 
We show that this limit on DM mass is notably more stringent than the unitarity ($\sim$100 TeV) and perturbativity limits (30-40 TeV) typically obtained for such models.
Furthermore, we find that for certain DM models, this completely rules out all the parameter space except a very narrow region near half of the Higgs mass, $M_h/2$ as discussed in Sec.~\ref{sec:dd}. While for other models, the surviving parameter space is constrained to below a few TeV and requires co-annihilation between DM and other dark sector particles.

This article is organized as follows. In Sec.~\ref{sec:higgs-dm-int}, we emphasize the importance of the Higgs interaction with DM in satisfying the observed relic density and enabling its detection. After that, in Sec.~\ref{sec:eft_loop}, we discuss the effective potential approach for computing one-loop corrections to the Higgs mass arising from various SM and BSM particles within simple scalar, vector, and fermionic DM models. Sec.~\ref{sec:relic} describes the methodology adopted for computing the relic density and Higgs mass corrections in this work. Additionally, we explore how the  DM relic abundance fixes the Higgs-DM coupling and the resulting Higgs mass corrections. We also discuss its implications for constraining the parameter space of some popular DM models. After that we discuss the direct detection prospects of DM candidates for these models in Sec.~\ref{sec:dd}. Here, we discuss the allowed DM parameter space once the limits from direct detection, colliders and constraints from DM-induced Higgs mass corrections are included. Finally, we summarize our findings and present our conclusions in Sec.~\ref{sec:conc}. The Lagrangians for scalar, vector, and fermionic DM, along with the corresponding WIMP-nucleon cross-section are provided in App.~\ref{appendixA}.


\section{ Role of Higgs-DM Interaction: DM Relic Density and Direct Detection}
\label{sec:higgs-dm-int}
%

\begin{figure}[!h]
    \centering
    \includegraphics[height=4.9cm]{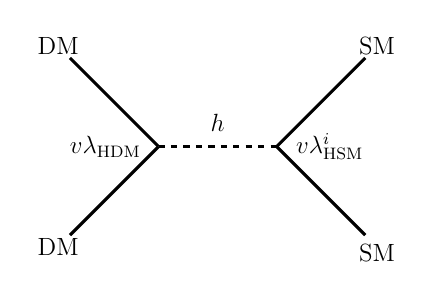}
    \caption{\centering DM annihilation via Higgs mediated channel.}
    \label{fig:DMDM_SMSM}
\end{figure}

\begin{figure}[!h]
    \centering
    \includegraphics[height=7.6cm]{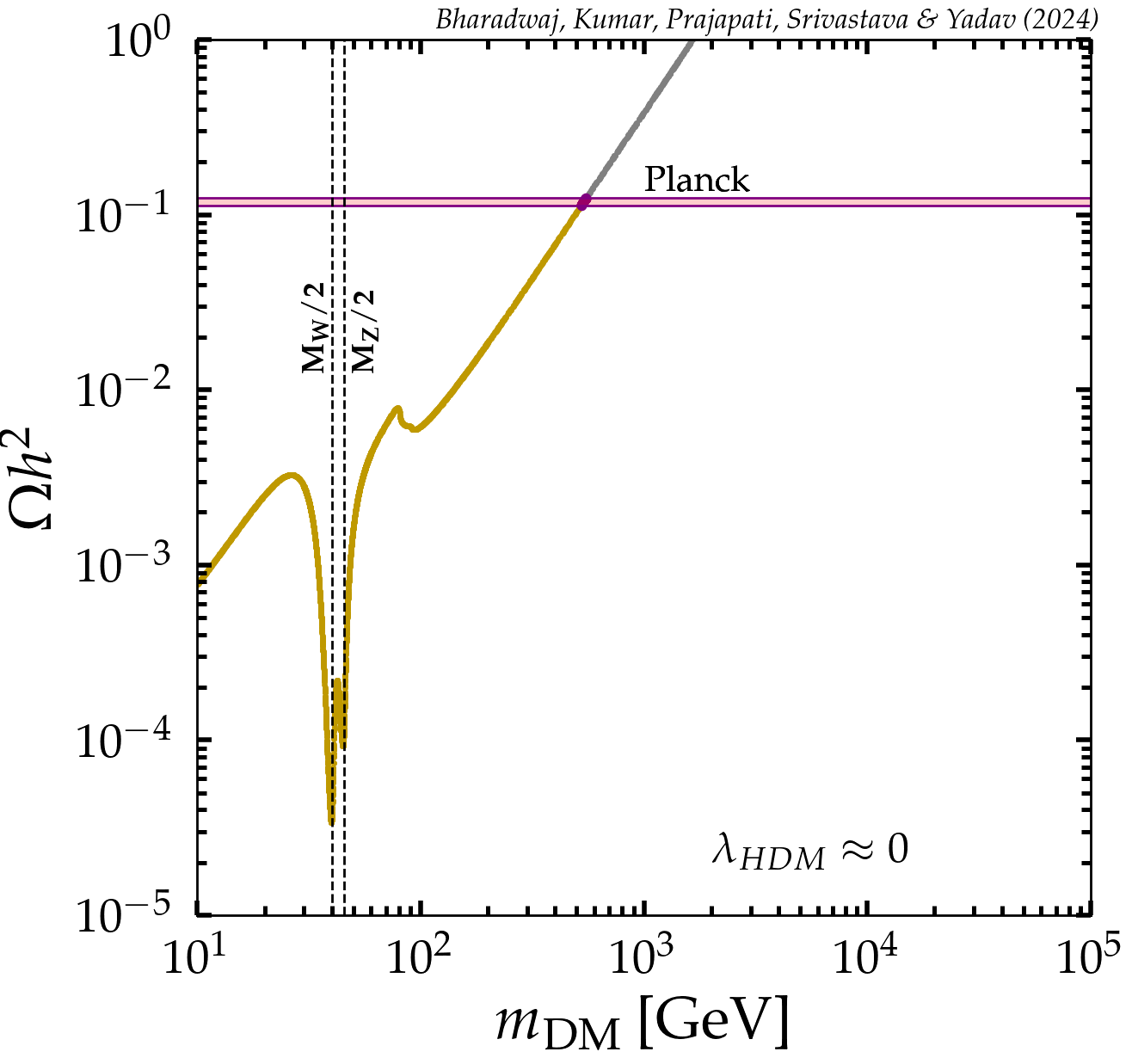}
     \includegraphics[height=7.6cm]{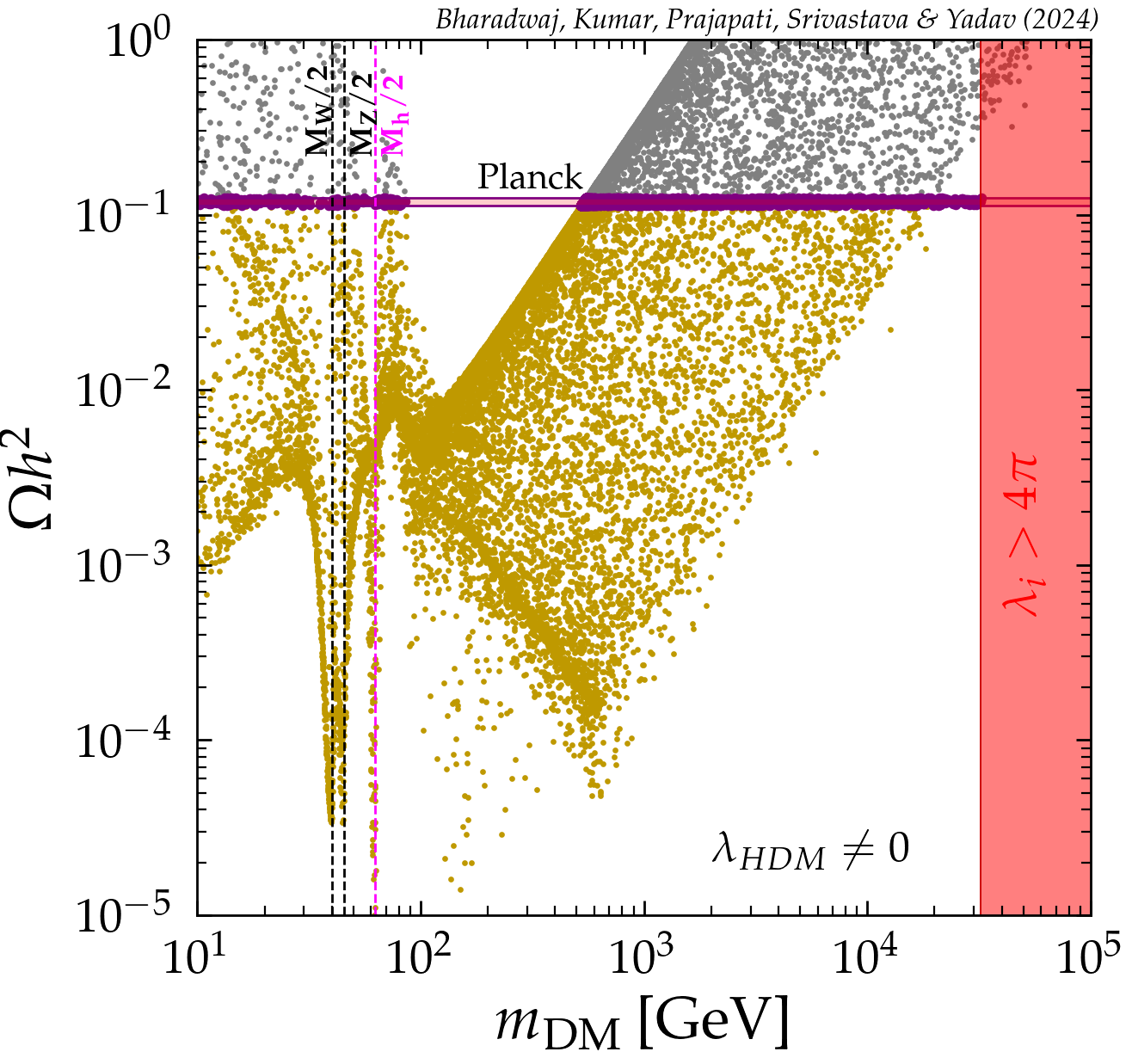}
    \caption{DM relic density as a function of DM mass. The left panel shows contributions exclusively from the W (co-annihilation)- and Z-mediated channels, while the right panel includes contributions from the Higgs-mediated channel in addition to the W (co-annihilation)- and Z-mediated channels. Yellow (gray) points represent DM under-abundant (over-abundant) regions, and purple points indicate points that satisfy the correct relic density. The red-shaded region is constrained by the naive perturbativity, see Sec.~\ref{Subsec:relic:IDM} for more details.}
    \label{fig:wzrelic}
\end{figure}

Let us begin by examining the crucial role of the Higgs-DM interaction plays in determining DM relic abundance. For a large class of models, the Higgs acts as a mediator connecting SM particles and DM, as shown in Fig.~\ref{fig:DMDM_SMSM}.
%
%
DM annihilation into SM particles as shown in Fig.~\ref{fig:DMDM_SMSM} is important to satisfy the correct observed relic density via freeze-out. 
This interaction depends on two key couplings: Higgs-SM ($\lambda^i_{\rm{H SM}}$) couplings, and Higgs-DM ($\lambda_{\rm{H DM}}$) coupling, with $\langle H \rangle = v/\sqrt{2}$ being the Vacuum Expectation Value (VEV) of the Higgs. Since the dominant Higgs-SM couplings are strongly constrained by experimental measurements, the DM annihilation is primarily controlled by the Higgs-DM coupling and DM mass. 
Note that, in addition to the Higgs boson, other particles, notably the SM Z boson, or even BSM particles, can also contribute to DM annihilation. While multiple channels may contribute to DM relic abundance, often in the high mass regime of DM, the Higgs-mediated channel becomes the dominant mechanism governing the DM relic abundance. For example, one such model is the inert doublet model (see, Sec. \ref{Subsec:relic:IDM}), where the DM relic density receives contributions not only from the Higgs-mediated channel but also from W (co-annihilation)- and Z-mediated channels. However, in the high mass regime, the Higgs-mediated channel provides the dominant contribution to the relic density as seen from Fig.~\ref{fig:wzrelic}.
The left panel illustrates only W (co-annihilation)- and Z-mediated channels, while the right panel shows the Higgs-mediated channel alongside the W- and Z-mediated contributions. 
From Fig.~\ref{fig:wzrelic}, one can infer that when there is no Higgs contribution, relic can be satisfied only in a narrow range as the coupling is fixed by gauge interaction strength. However, once we include the Higgs-DM interactions as well, then the DM can have correct relic density over a much wider mass range. Notably, when the DM mass is significantly larger than the W and Z masses, the Higgs-mediated channel becomes the dominant contributor to the relic density, as shown in the right panel of Fig.~\ref{fig:wzrelic}. 
Hence, Higgs-DM interaction is crucial for satisfying relic density over a wide range of DM mass. Note that an upper limit on this DM mass range arises from unitarity considerations, which restrict the WIMP mass up to approximately 100 TeV \cite{Griest:1989wd,Smirnov:2019ngs}. Additionally, naive perturbativity of the Higgs-DM coupling constrains DM mass around ($30-40$) TeV\footnote{The more accurate S-matrix computation of perturbativity typically allows for a more relaxed upper limit on Higgs-DM coupling, see \cite{Goodsell_2018}.}.

\begin{figure}[!h]
\centering
\includegraphics[width=.5\textwidth]{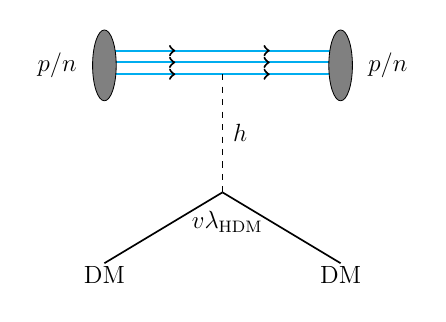}
\caption{\centering Diagram responsible for the direct detection of the DM.}
\label{dd}
\end{figure}

The Higgs-DM coupling not only plays a pivotal role in determining the relic abundance of DM but also in its detection prospects. It acts as a mediator connecting DM to heavy target nuclei like Xenon (Xe) used in direct detection experiments as shown in Fig.~\ref{dd}.
%
 These experiments put limits on the WIMP-nucleus scattering cross-section by measuring the nuclear recoil resulting from WIMP-nucleus collisions.
\begin{figure}[!h]
    \centering
    \includegraphics[width=0.85\linewidth]{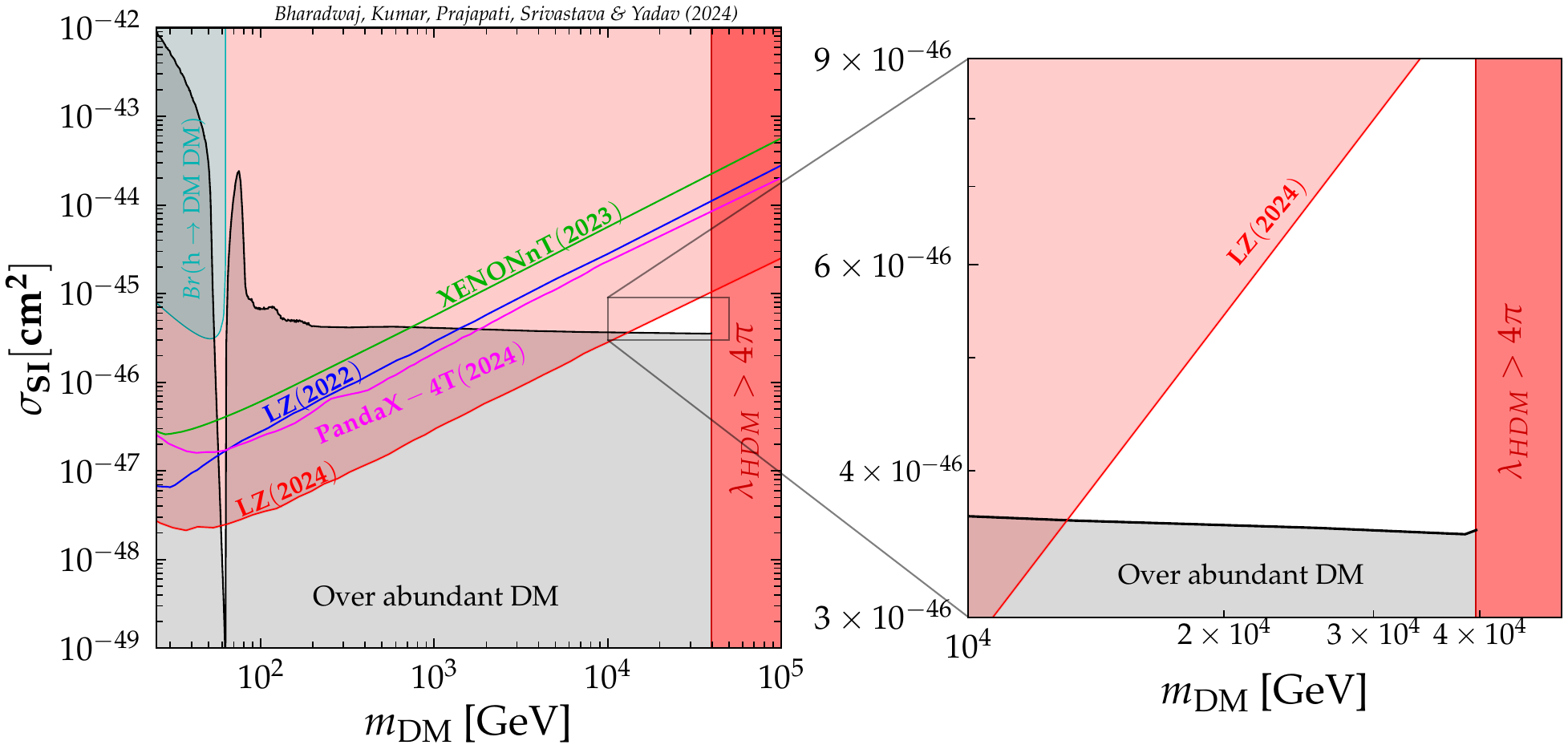}~~~
    \caption{WIMP-nucleon scattering cross-section as a function of DM mass for real singlet scalar DM, see Sec.~\ref{Subsec:relic:RSM} and Sec.~\ref{Subsec:dd:RSM} for more details. When DM is heavier than $\sim40$ TeV,  the Higgs-DM coupling $\lambda_{HS}$ becomes non-perturbative. Various direct detection experimental constraints are also shown.}
    \label{fig:purturb}
\end{figure}
Current direct detection experiments, such as XENONnT~\cite{XENON:2023cxc}, PandaX-4T~\cite{PandaX:2024qfu}, and LUX-ZEPLIN (LZ)~\cite{LZ:2022lsv,LZCollaboration:2024lux} put stringent limits on DM within the mass range of approx 1 GeV to 10 TeV, see Sec.~\ref{sec:dd} for more details.
This mass range is determined by experimental sensitivity, as lower mass DM imparts smaller recoil energies to the nucleus, making detection challenging. Conversely, higher mass DM produces more energetic recoils, which are easier to detect. However, the expected DM flux and hence interaction rate decreases appreciably with increasing mass, requiring larger and larger detector volumes and extended exposure times to accumulate a sufficient number of events. Consequently, a vast parameter space for heavy DM masses remains poorly constrained by current generation of direct detection experiments. For example, consider the case of the real singlet scalar model described in Eq.~\eqref{eq:eftsdm}. In this specific case, Fig.~\ref{fig:purturb} illustrates that after accounting for tree-level perturbative limit $\lambda_{\rm{HDM}} < 4 \pi$, current direct detection experiments allow the DM mass to be roughly in the range of $10$ - $40$ TeV, see Sec. \ref{sec:dd} for details. Closing this allowed parameter space poses a significant challenge for current generation of such experiments. Detecting or completely ruling out DM candidates with masses around 40 TeV would require detectors of much larger size and long exposure time. The proposed future experiments such as PandaX-xT \cite{PANDA-X:2024dlo} and XLZD \cite{XLZD:2024nsu} will hopefully be able to put more stringent limits on this heavy DM mass regime.

However, an often overlooked factor in constraining this parameter space is the impact of heavy DM on corrections to the Higgs mass.
In this work, we demonstrate that, for a broad class of DM models, the heavy DM mass regime can be excluded due to DM-induced large loop corrections to the Higgs mass. In this regime, the DM-induced loop corrections become so significant that the Higgs mass cannot be maintained at its experimentally observed value of $125.20 \pm 0.11 $ GeV \cite{ParticleDataGroup:2024cfk}.
 This implies that for any DM model where Higgs-DM coupling is required to satisfy relic abundance for DM mass range of TeV or above, there exists an upper bound on DM mass, typically around few TeVs. 
Furthermore, we show that by incorporating Higgs mass corrections, along with the constraints from DM phenomenology, current direct detection experiments can already completely rule out some of these models, except in a fine-tuned narrow mass range near $M_h/2$.

\section{Effective potential and estimate of DM-induced Higgs mass correction}
\label{sec:eft_loop}

As discussed before, in the high mass region of DM, specifically when ($M_{\rm{DM}} \gsim \mathcal{O}(1)$ TeV), the Higgs-mediated channel plays a substantial role in satisfying the relic density of DM. This motivates our focus on scenarios where DM interacts with the visible sector via the Higgs. We begin by examining the simplest cases of scalar, vector, and fermionic DM candidates. To highlight key features of Higgs-DM interactions, we select a simple model for each type, where DM interacts with SM fields through the Higgs. The relevant terms in  Lagrangian density corresponding to these scenarios are:
\begin{align}
-{\cal L}_{\rm Scalar} & \supset
\mu_S^2\, S^2 + \frac{\lambda_{HS}}{2}\,H^\dagger H \,S^2 +\frac{\lambda_S}{2}\, S^4\,,
\label{eq:eftsdm}
\\
-{\cal L}_{\rm Vector} & \supset 
\frac{1}{2}\, \mu_V^2 V_\mu V^\mu + \frac{1}{4}\, \lambda_{HV} H^\dagger H {V_\mu V^\mu}
+ \frac{1}{4} \lambda_V (V_\mu V^\mu )^2 
\label{eq:eftVec}\,,
\\
-{\cal L}_{\rm Fermion} & \supset \mu_\psi \overline{\psi} \psi 
+ \,\frac{\lambda_{H\psi}}{\Lambda} H^\dagger H ~\overline{\psi} \psi
\label{eq:eftFer}\,.
\end{align}
Here, $H$ is the SM Higgs doublet, $S, \ V, \ \text{and} \ \psi$ are the scalar, vector, and fermionic DM, and the Higgs-DM coupling
is denoted by $\lambda_{HS}, \ \lambda_{HV}, \ \text{and} \  \lambda_{H\psi}$, respectively.  

The scalar DM scenario is represented in Eq.~\eqref{eq:eftsdm}, where  $S$ is the real scalar field. Here, we have also assumed a $Z_2$ symmetry under which all SM particles are even while $S$ is odd. This allows us to forbid the tree-level cubic term $H^\dagger H S$ which would lead to $S \to h h$ decay. We further assume that $S$ does not get any VEV i.e. $\langle S \rangle = 0$, hence $Z_{2}$ remains unbroken. The stability of this DM candidate is ensured by $Z_{2}$ symmetry. The effective vector DM case is shown in Eq.~\eqref{eq:eftVec}. Here, as well, the stability of DM is ensured by the imposition of a $Z_{2}$ symmetry, similar to the case of scalar DM. Note that Eq.~\eqref{eq:eftVec} is only an effective model, and its UV-completion requires $V$ to be a gauge boson of a new gauge symmetry. One such UV completion is the dark abelian SM gauge extension, where the  $ Z_{2} $ symmetry is also naturally present, providing inherent stability to the vector DM candidate \cite{Lebedev:2011iq}. 
Finally, Eq.~\eqref{eq:eftFer} describes the effective fermionic DM scenario. It is worth noting that in this case also, the Higgs–DM coupling is non-renormalizable. However, it also can be easily UV-completed through appropriate extensions \cite{PhysRevD.82.055026,Djouadi:2011aa,Arcadi:2019lka}.

\begin{figure}[!h]
    \centering
    \includegraphics[width=0.3\linewidth]{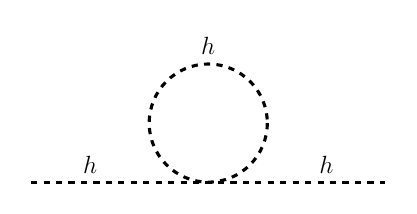}
    \includegraphics[width=0.3\linewidth]{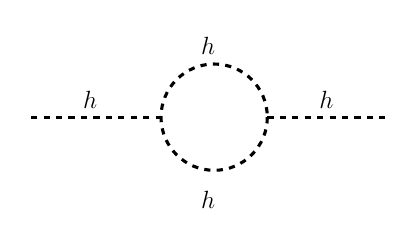}
    \includegraphics[width=0.3\linewidth]{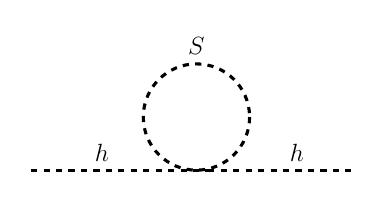}
    \includegraphics[width=0.3\linewidth]{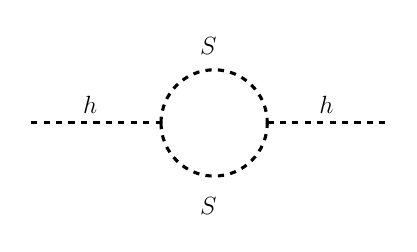}
    \includegraphics[width=0.3\linewidth]{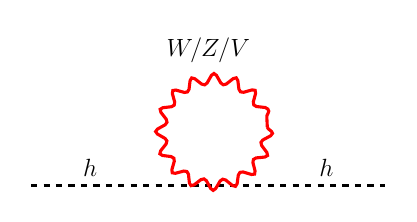}
    \includegraphics[width=0.3\linewidth]{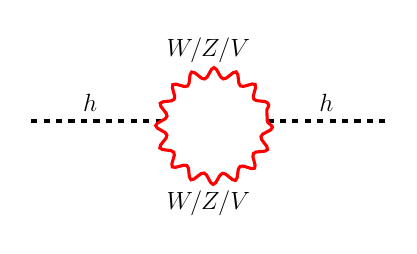}
    \includegraphics[width=0.3\linewidth]{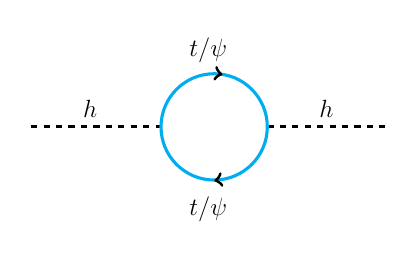}
    \caption{One-loop corrections to Higgs mass due to SM particles and DM. Note that for a given DM case, only the BSM loops relevant to that model will be present.}
    \label{Fig:Loop_diagrams}
\end{figure}
As previously mentioned, loop corrections to the Higgs mass induced by DM particles play a crucial role in determining the viable parameter space for DM mass. Therefore, we first examine the one-loop corrections to the Higgs mass arising from different SM and BSM particles as shown in Fig.~\ref{Fig:Loop_diagrams}. 
We follow the Coleman-Weinberg effective potential approach~\cite{Coleman:1973jx}. In the $\overline{\rm MS}$ renormalization scheme, the one-loop corrections to the Higgs effective potential are obtained as~\cite{Coleman:1973jx, Sher:1988mj}
\begin{equation}
\label{eq:veff}
V_{\rm{eff}}(h_{c})=V_0(h_{c})+\frac{1}{64\pi^2}\sum_{i} n_i
m_i^4(h_{c})\left[\log\frac{m_i^2(h_{c})}{\mu^2}-\rm{c}_i\right]. 
\end{equation}
Here, $h_{c}$ is the real constant background field, $n_i$ are degrees of freedom, $m_i$ are field-dependent masses of particle which contribute to the one-loop self-energy, and $\rm{c}_i$ are constants given by, 
\begin{itemize}
\centering
    \item[] For vector bosons, \quad $\rm{c}_i=\frac{5}{6}$, 
    \item[] For other particles, \quad $\rm{c}_i=\frac{3}{2}$.
\end{itemize}
The first term in Eq.~\eqref{eq:veff}, $V_0 (\rm{h_{c}})$ is the tree-level potential in terms of background field, and the second term denotes the one-loop effective potential considering all other particles which interact with Higgs in a particular model.  

Let's first discuss the one-loop correction to Higgs mass in SM. The tree-level potential of the SM Higgs is given as,
 \begin{equation}\label{Eq:Higgs_Potentail}
     (V)_{SM} = \mu_{H}^{2}H^{\dagger}H +\frac{\lambda_{H}}{2}(H^{\dagger}H)^{2}\,,
 \end{equation}
where $\lambda_H$ is the Higgs self-quartic coupling. 
The $SU(2)_L$ doublet $H$  after spontaneous electroweak symmetry breaking, can be written as
 \begin{eqnarray} \label{Eq:Higgs_Doublet_vev}
     H = \frac{1}{\sqrt{2}}\begin{pmatrix}
         \sqrt{2}G^{+}\\
        v + h + i G^{0}
     \end{pmatrix}.
 \end{eqnarray}
Here $h$ is the SM Higgs and $G^{+}$/$G^{0}$ are the charged/neutral Goldstones. After electroweak symmetry breaking, the Goldstones get absorbed by the $W^\pm, Z$ gauge bosons, which become massive. The remaining physical scalar ``Higgs" gets mass which at tree level is given by $m^2_h = \lambda_H v^2$. Theoretically, the Higgs self-quartic coupling $\lambda_H$ can take any value in the range $\lambda_H  \in [0, 4 \pi]$. The lower limit coming from the boundedness from below and upper limit from ``naive perturbativity" requirements. This range of values of $\lambda_H$ implies that $m_h$ can also vary over a corresponding range. In rest of this work we assume this theoretical range for $m_h = \lambda_H v^2$; $\lambda_H  \in [0, 4 \pi]$  while computing loop corrections to the Higgs mass. Note that it is the loop corrected Higgs mass $M_h$ that should match the experimentally measured values of $125.20 \pm 0.11$ GeV \cite{ParticleDataGroup:2024cfk}\footnote{Strictly speaking the experimentally measured Higgs mass corresponds to the ``on-shell mass" which is not exactly same as the loop corrected mass. However, throughout this work we ignore this small difference as it has negligible effect on our results.}. Furthermore, all the loop computations in this work will be done exclusively in the $\overline{\rm{MS}}$ scheme.    

Now, let's discuss the possible loop corrections to Higgs mass within SM. To derive the effective potential as expressed in Eq.\eqref{eq:veff}, we first use the Higgs potential given in Eq.\eqref{Eq:Higgs_Potentail} to obtain the field-dependent tree-level potential in the SM,
\begin{eqnarray}
(V_{0})_{SM} (h_{c}) = \frac{\mu_{H}^{2}}{2}h_{c}^{2} +  \frac{\lambda_{H}}{8}h_{c}^{4} .
\label{eq:tree-v0}
\end{eqnarray}
with the field dependent expansion of the $SU(2)_L$ doublet H given by 
\begin{eqnarray} \label{Eq:Higgs_Doublet}
     H = \frac{1}{\sqrt{2}}\begin{pmatrix}
         \sqrt{2}G^{+}\\
        h_c + h + i G^{0}
     \end{pmatrix}.
 \end{eqnarray}
The second term of Eq. \eqref{eq:veff} gets contribution from all the particles that couple to SM Higgs. This includes SM fermions, particularly top quark due to its large mass, gauge bosons (W, Z), Goldstones, as well as self-corrections due to Higgs self-quartic-coupling. Their field-dependent masses are given as, 
\begin{equation}
\begin{split}
   & m_{h}^{2}(h_{c}) = \mu_{H}^{2} + \frac{3}{2}\lambda_{H}h_{c}^{2},~~~ m_{G^{0}}^{2}(h_{c})=m_{G^{\pm}}^{2}(h_{c}) = \mu_{H}^{2} + \frac{\lambda_{H}h_{c}^{2}}{2}, ~~~m_{W}^{2}(h_{c}) = \frac{g^{2}}{4}h_{c}^{2}, \\
    & m_{Z}^{2}(h_{c}) = \frac{(g^{2}+g'^{2})}{4}h_{c}^{2}, ~~~m_{t}^{2}(h_{c}) = \frac{Y_{t}}{\sqrt{2}}h_{c} \ .
    \end{split}
\end{equation}
Here, $g'$, and $g$ denote the gauge couplings  for $U(1)_{Y}$ and $SU(2)_{L}$ respectively, and $Y_{t}$ is top quark Yukawa coupling. The last ingredient is the degree of freedom, which is given as, $n_{i} = \{ 1,1,2,6,3,-12 \}$ for species $i= \{h,G^{0},G^{\pm},W,Z,t\} $. 
Considering that the effective potential, $V_{\rm{eff}}(h_{c})$ has an extremum for $ h_{c} = v = 246.22$ GeV, the Higgs mass in SM at one-loop can be computed as~\cite{Casas:1994us,Hambye:2007vf}
\begin{equation}
  (M_{\text{h}}^2)_{\text{SM}} = \frac{d^2 (V_{\rm{eff})_{\text{SM}}}}{d h_{c}^2} \Big|_{ h_{c} =v} = m^2_{\rm{h}} + (\delta_{1}m^2_{\rm{h}})_{\mathrm{SM}}\,, 
\end{equation} 
with
 \begin{align}\label{Eq:SM_Hggs_mass_correction}
   (\delta_{1}m^2_{\rm{h}})_{\rm{SM}} = & \frac{1}{32 \pi^2} \Big[ 3 \lambda_{H} f\left( m_h^2 \right)+ 9(\lambda_{H} v)^{2}\log\frac{m_{h}^{2}}{\mu^{2}} + \lambda_{H} f\left( m_{G^{0}}^{2} \right)+ (\lambda_{H} v)^{2}\log\frac{m_{G^{0}}^{2}}{\mu^{2}}    \nonumber \\ & + 2\lambda_{H} f\left( m_{G^{+}}^{2} \right)+ 2(\lambda_{H} v)^{2}\log\frac{m_{G^{+}}^{2}}{\mu^{2}}  -36 Y_{t}^{2} f(m_{t}^{2}) -12 Y_{t}^{4}v^{2}   \nonumber \\ & + \frac{9 (g^{2}+g'^{2})f(m_{Z}^{2})}{2} + \frac{3 (g^{2}+g'^{2})^{2}v^{2}}{2} + 9 g^{2} f(m_{W}^{2}) + 3g^{4}v^{2}  \Big].
 \end{align}
 Where $f(x^2)=x^2 \left(\log \frac{x^2}{\mu^2}-1\right)$, $m^2_{\rm{h}} =  \lambda_{H}\, v^2$, and $\mu$ is the renormalization scale. Note that throughout this work we use the small letter $m_i$ to denote the tree-level mass of a given particle ``i" and the capital letter $M_i$ to denote the loop corrected mass of the same ``i" particle.

For BSM scenarios, where only SM Higgs  acquires VEV, the same framework could be extended, and the one-loop corrected Higgs mass is given as follows,
\begin{equation}
    M_{\rm{h}}^{2} = (M_{\text{h}}^2)_{\text{SM}} + (\delta_{1} m_{\text{h}}^2)_{\text{BSM}} \ .
\end{equation}
For all the different types of Higgs-DM interactions written in Eqs.~(\ref{eq:eftsdm}-\ref{eq:eftFer}), the new contributions are provided by DM running in the loop, see Fig. \ref{Fig:Loop_diagrams}. Further note that from Eq.~\eqref{Eq:SM_Hggs_mass_correction}, it is clear that bosons and fermions give opposite contributions to the Higgs mass. Hence, in BSM scenarios, correction to Higgs mass by bosonic and fermionic DM needed to be discussed separately. We will first start with the bosonic DM case.


\subsection{Correction to the Higgs mass for bosonic DM}
\label{subsec:loop-bosonic}
 
First, we consider the scalar and vector boson DM cases. We take the models listed in Eqs.~(\ref{eq:eftsdm}-\ref{eq:eftVec}) as example models but similar constraints are applicable to many non-SUSY bosonic DM models. Basically, any bosonic DM where Higgs-DM interaction plays a role in obtaining the correct DM relic density will have similar constraints. 
Coming back to the scalar and vector DM models of Eqs.~(\ref{eq:eftsdm}-\ref{eq:eftVec}) and  using Eq.~\eqref{eq:veff}, one-loop contributions to the Higgs mass in these cases can be written as,
\begin{equation}
    \begin{split}
        &(\delta_{1} m_{\text{h}}^2)_{\text{Scalar}} = \frac{1}{32 \pi^{2}} \left[  \lambda_{HS}f(m_{S}^{2}) + (v \lambda_{HS})^{2} \log \frac{m_{S}^{2}}{\mu^{2}}  \right]\,, \\
        &(\delta_{1} m_{\text{h}}^2)_{\text{Vector}} = \frac{\lambda_{HV}}{128 \pi^{2}} \left[ -2 m_{V}^{2} +2 v^{2} \lambda_{HV} + 3 (2m_{V}^{2} + v^{2} \lambda_{HV}) \log \frac{m_{V}^{2}}{\mu^{2}} \right]\,.
    \end{split}
\end{equation}
where,  $m_{S}^{2} = \mu_{S}^{2} + \frac{\lambda_{SH}}{2}v^{2}$, $m_{V}^{2} = \mu_{V}^{2} + \frac{\lambda_{HV}}{4}v^{2}$. The Higgs-DM coupling $\lambda_{\rm{HDM}} \equiv \lambda_{\rm{HS}}$ and $\lambda_{\rm{HDM}} \equiv \lambda_{\rm{HV}}$ for the scalar and vector DM, respectively.

\begin{figure*}[h]
\includegraphics[width=0.45\linewidth]{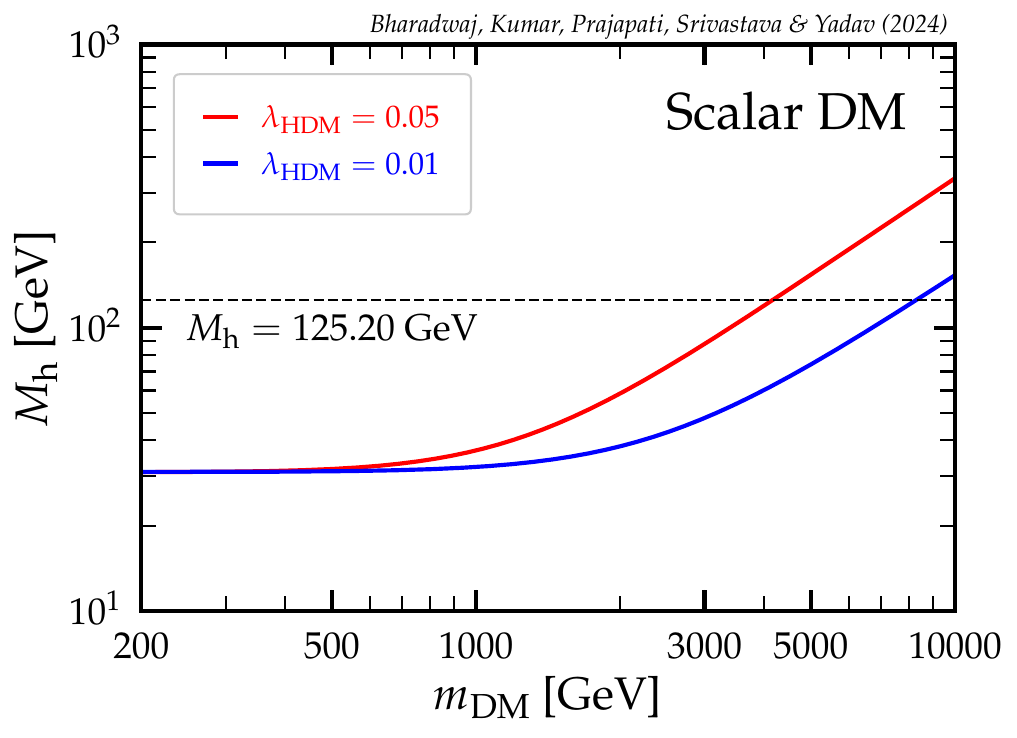}
\includegraphics[width=0.45\linewidth]{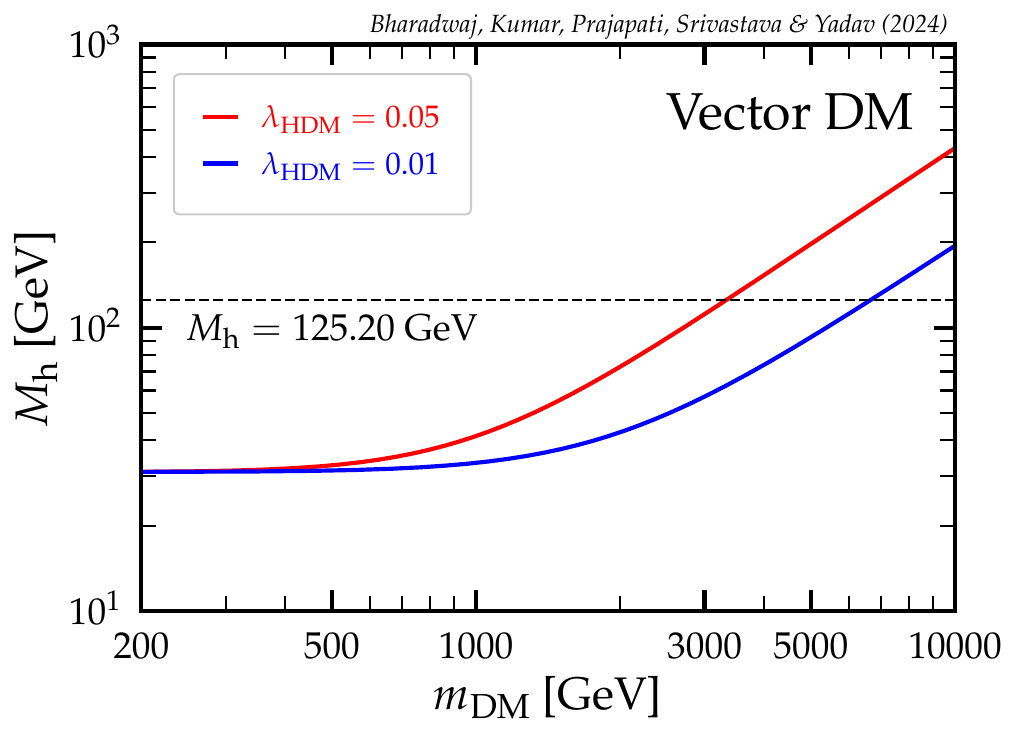}
    \caption{ One-loop corrected Higgs mass $M_{\rm{h}}$ vs mass of bosonic DM. \textbf{Left:} For the scalar DM case, \textbf{Right:} For the vector boson DM. The blue and red line corresponds to two different values of Higgs-DM coupling.
    In making these plots, we have taken the conformal limit of tree level Higgs mass $m_h \to 0$. See text for more details.}
    \label{Fig:Scalar_Boson_DM_Hmass_correc}
\end{figure*}

In Fig. \ref{Fig:Scalar_Boson_DM_Hmass_correc}, we plotted one loop corrected Higgs mass as a function of DM mass, the left panel shows the case for scalar DM, and the right panel shows vector boson DM. From Fig.~\ref{Fig:Scalar_Boson_DM_Hmass_correc}, one can infer that the loop contribution to Higgs mass is positive for bosonic DM.
In making Fig.~\ref{Fig:Scalar_Boson_DM_Hmass_correc}, we have taken the scale to be fixed at top mass, i.e. $\mu=m_{t}$. In principle, variations in the renormalization scale are compensated by the scale dependence of the running quartic couplings. However, for simplicity, we currently disregard the running of couplings and fix the scale at the top quark mass. We will revisit this problem and will perform a more detailed two-loop computation in Sec.~\ref{sec:relic}, taking care of the above shortcomings.

Note that for a given Higgs-DM coupling,  the loop correction increases with increasing DM mass.
Therefore, to maintain the loop corrected Higgs mass $M_h$ at 125 GeV, one has to adjust the Higgs self-quartic coupling $\lambda_H$ appropriately\footnote{Usually in order to keep the BSM induced loop corrections under control, the BSM-Higgs coupling is fine-tuned. However, as we discussed and will elaborate further in Sec.~\ref{sec:relic}, the strength of Higgs-DM coupling is decided by the requirement that DM should have correct observed relic abundance, and hence, this coupling cannot be fine-tuned beyond a point.}, a freedom that is allowed since $\lambda_H$ is not yet measured.
Therefore, as the DM-induced loop correction increases, one can correspondingly decrease  $\lambda_H$ and hence the tree level mass $m_h$ so that $M_h$ remains fixed at 125 GeV. Beyond a certain DM mass, the DM-induced loop corrections become so large that $M_h$ cannot be kept at 125 GeV even when we take the conformal limit of $m_h \to 0$. 

In order to further elaborate this point, in plotting Fig.~\ref{Fig:Scalar_Boson_DM_Hmass_correc} we have taken the conformal limit of $m_h \to 0$; this also makes the one loop Higgs self-correction to be effectively zero. Then for benchmark values of the Higgs-DM coupling, we have computed the one-loop corrected Higgs mass ($M_{\rm{h}}$) as shown in Fig.~\ref{Fig:Scalar_Boson_DM_Hmass_correc}, where the blue and red lines represent two distinct values of the Higgs-DM quartic coupling.
The black dashed line represents the experimentally observed best fit value of Higgs mass, $M_{\rm{h}} = 125.20$ GeV.  Thus, if DM mass is beyond a certain point, the DM-induced loop corrections are too large, and $M_h$ cannot be kept at its measured value even in the $m_h \to 0$ limit. This puts a strong upper limit on the DM mass. As mentioned before, similar limits can be placed on all non-SUSY bosonic DM scenarios whenever the Higgs-DM coupling can be constrained by DM relic abundance. We will revisit this in more elaborate detail in Sec.~\ref{sec:relic}.
%


\subsection{Correction to the Higgs mass for fermionic DM}
\label{subsec:loop-fermionic}

We now turn our attention to the loop corrections to Higgs mass induced by fermionic DM. Again, the analysis presented here is applicable to any fermionic DM scenario where the Higgs-DM coupling,  at least in the high DM mass regime, can be constrained by the DM relic density.  
For the fermionic DM scenario given in Eq. \eqref{eq:eftFer}, the one-loop correction to Higgs mass is given as,
\begin{equation}\label{eq:FermionDM}
    \begin{split}
          &(\delta_{1} m_{\text{h}}^2)_{\text{Fermion}}= -\frac{\lambda_{H \psi}}{\Lambda} \frac{m_{\psi}^{2}}{4 \pi^{2}}\left[ 
          m_{\psi} \left(  \log \frac{m_{\psi}^{2}}{\mu^{2}} -1   \right)+ \frac{\lambda_{H \psi}}{\Lambda}v^{2}
          \left(  3 \log \frac{m_{\psi}^{2}}{\mu^{2}} -1\right)   \right] \,,
    \end{split}
\end{equation}
where, $m_{\psi} = \mu_{\psi} +\frac{\lambda_{H\psi}}{\Lambda}\frac{v^2}{2}$, and $\Lambda$ is the cut-off scale beyond which a full Ultra-Violet (UV) complete theory is needed. Note that the Higgs-DM coupling is non-renormalizable. This means that the number of counter-terms needs to cancel divergences, keeps on increasing indefinitely as we go to higher and higher loops.
However, the infinities can still be absorbed loop by loop by adding appropriate counter terms and computations for loop corrections up to a given loop can still be performed~\cite{Sher:1988mj}.

From Eq. \ref{eq:FermionDM}, it is clear that the fermionic DM contribution in the one-loop correction to the Higgs mass is negative, which leads to a decrease in the loop corrected Higgs mass $M_h$ from its tree-level value $m_h$. However, as before the Higgs self-quartic coupling $\lambda_H$ is a free parameter. Therefore, the negative loop correction can countered by appropriately increasing the tree-level Higgs mass $m_h$  by enhancing this coupling. Thus, as the DM mass and hence DM-induced loop corrections increase, one needs to correspondingly increase $\lambda_H$ so that the loop corrected mass remains fixed at $125$ GeV\footnote{As before, here also the requirement of obtaining correct observed DM relic abundance takes away the freedom to fine-tune the Higgs-DM coupling beyond a point.}. Again, this process cannot be continued indefinitely as $\lambda_H$ is constrained from above by perturbativity and unitarity limits. For example, taking $\lambda_H \approx \sqrt{4\pi}$ gives an upper limit on tree-level Higgs mass $m_h \approx 464$ GeV. Using the more careful limit derived from partial-wave unitarity for the two-body scattering of gauge bosons gives an upper limit of $m_h \approx 712$ GeV~\cite{Lee:1977eg,Logan:2022uus}. Thus, when the DM-induced loop corrections become large enough then the loop corrected Higgs mass $M_h$ can no longer be kept at $125$ GeV. This again provides an upper limit on the DM mass.

\begin{figure}[h]
    \centering
    \includegraphics[width=0.55\linewidth]{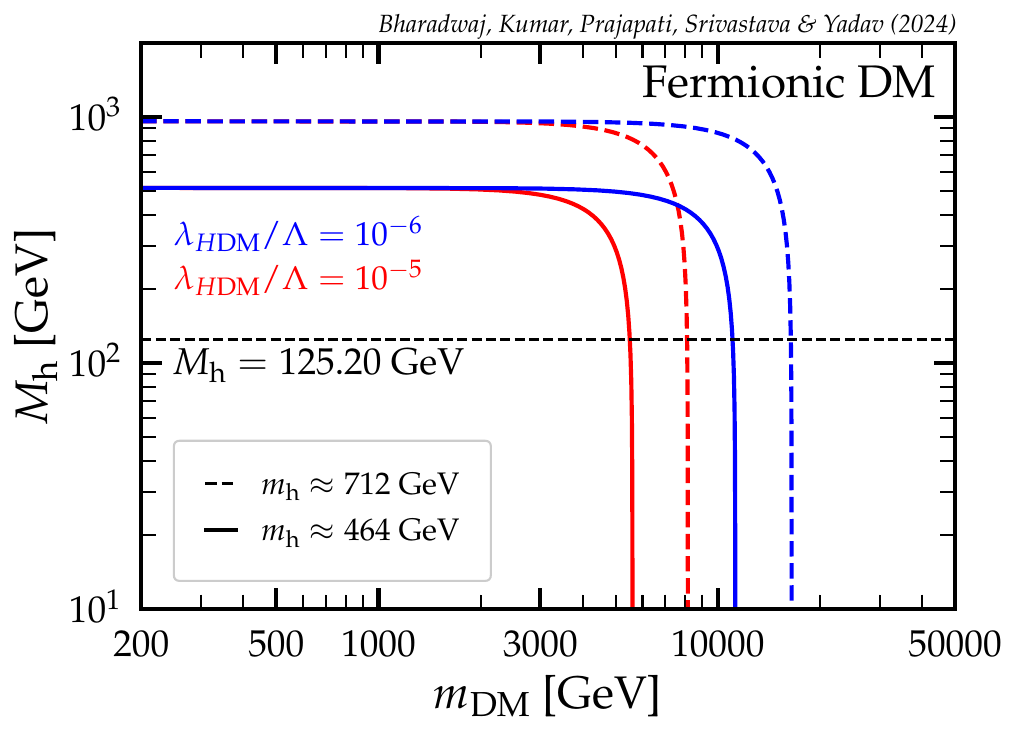}
    \caption{ One-loop corrected Higgs mass for the fermionic DM case. The blue and red lines correspond to two different values of Higgs-DM coupling. Solid and dashed lines correspond to tree-level Higgs masses of $m_h \approx 464$ GeV and $m_h \approx 712$ GeV, respectively. See text for details.}
    \label{Fig:FermionDM_H_correc}
\end{figure}

In Fig.~\ref{Fig:FermionDM_H_correc}, we have plotted loop corrected Higgs mass $M_h$ as a function of fermionic DM mass. We have taken two benchmark values of tree-level Higgs mass. 
The first benchmark (dashed line) corresponds to the upper bound on  tree level Higgs mass ($\approx 712$ GeV) derived by partial-wave unitarity for the two-body scattering of gauge bosons \cite{Lee:1977eg,Logan:2022uus}, while the second benchmark (solid line) is for $\lambda_{H} \approx \sqrt{4 \pi}$. The one-loop corrected Higgs mass is shown in red and blue color for the two different values of Higgs-DM coupling. Again, we fix the scale at the top quark mass.

From Fig.~\ref{Fig:FermionDM_H_correc}, one can infer that up to a particular value of DM mass, there is only a slight correction in Higgs mass which remains almost a flat line. This nearly flat line is dominated by Higgs self and top corrections. However, once the DM contribution becomes dominant,  due to the large negative correction by the fermionic DM, the one-loop corrected Higgs mass rapidly decreases and eventually approaches zero. Hence, for a given Higgs-DM coupling, there exists an upper bound on the fermionic DM mass beyond which the Higgs mass cannot be maintained at its experimentally observed value. This can restrict the allowed parameter space of fermionic DM significantly.
 A more detailed analysis of possible UV completions of fermionic DM models will be studied in the follow-up works.

One shortcoming of the above analysis is that we have not considered the renormalization group evolution (RGE)  of the couplings. However, to make the above analysis more precise, one needs to take the RGE of couplings into account and use an improved effective potential. In the next sections Sec.~\ref{sec:relic} and Sec.~\ref{sec:dd} we improve our current analysis for three different models of scalar DM. 
In order to take the RGE improved effective potential into account in our analysis, we compute the Higgs mass correction using the SARAH ~\cite{Staub:2013tta, Staub:2015kfa} and SPheno ~\cite{Porod:2003um, Porod:2011nf} packages that allow precise higher-order calculations. Additionally,  micrOMEGAs software package~\cite{Belanger:2014vza} has been employed which efficiently computes DM observables, including DM relic density and DM-nucleon scattering cross-section.
In SARAH and SPheno the correction to the Higgs mass is computed via a two-scale matching procedure, as outlined in Ref.~\cite{Staub:2017jnp} for SUSY models. This procedure has also been adapted to non-SUSY models by excluding the $\overline{\rm{MS}}$ to $\overline{\rm{DR}}$ parameter translation and ignoring SUSY threshold effects as explained in~\cite{Staub:2013tta}. Consequently, all self-energy corrections are computed in the $\overline{\rm{MS}}$ scheme instead of $\overline{\rm{DR}}$. 
The SARAH + SPheno framework enables us to precisely evaluate two-loop Higgs mass corrections while ensuring that SM RGEs are taken into account up to the three loop-level. Notably, the dependence on the renormalization scale $\mu$ has only a limited effect on the two-loop corrections to the Higgs mass, as discussed in Ref.~\cite{Braathen:2017izn}. We rely on the default top mass scale provided by SPheno for consistency across the calculations. This procedure offers a reliable approach to studying Higgs mass corrections and other related quantities and is employed henceforth throughout the rest of this paper.

\FloatBarrier
\section{Interplay between DM Relic Density and DM-induced Higgs Mass Correction} \label{sec:relic}
In the early universe, particles in the dark sector maintain thermal equilibrium with SM particles due to a balance between production and annihilation processes. As the universe expanded and cooled, the temperature eventually dropped below the mass of dark sector particles. At this stage, thermal production ceased for these particles, but their annihilation and decay processes persisted, leading to their gradual departure from the thermal bath. However, the lightest dark sector particle, which is stable, remained unaffected by decay. As the temperature decreases further, this stable particle eventually decouples from the thermal bath, leading to a ``freeze-out" of its relic density. This relic density can be calculated and compared with observational data, such as those from the Planck satellite~\cite{Planck:2018vyg}. According to Planck measurements, the current observed DM relic abundance is constrained within the range 
$0.1126 \leq \Omega h^2 \leq 0.1246$ at the $3\sigma$ confidence level.

In this section, for some popular DM models,  we explicitly compute the relic density and discuss the interplay between it and DM-induced loop corrections to the Higgs mass. We show that as the DM mass increases, one has to increase the Higgs-DM coupling in order to satisfy the correct relic density, for example, see the right panel of Fig.~\ref{fig:SSDM_RDtree}. Thus, relic density constrains the strength of Higgs-DM coupling from below.
Now, as discussed in the previous section, the DM mass and Higgs-DM coupling control the magnitude of DM-induced loop corrections to the Higgs mass, which increases with increasing DM mass and Higgs-DM coupling. Thus, there comes a point beyond which the DM-induced loop corrections are so large that the loop corrected Higgs mass cannot be maintained within the experimentally measured value\footnote{As discussed before, we ignore the small difference between two-loop Higgs mass obtained using $\overline{\rm{MS}}$ scheme and the pole mass. This small difference has a negligible effect on our results. } of Higgs mass, $M_H = 125.20\pm0.11$ GeV~\cite{ParticleDataGroup:2024cfk}, even in the conformal limit of $m_h \to 0$. This puts a strong limit on the maximum allowed DM mass which turns out to be few TeVs. We implement this program, using SARAH~\cite{Staub:2013tta, Staub:2015kfa} and SPheno~\cite{Porod:2003um, Porod:2011nf} for computing loop corrections and micrOMEGAs~\cite{Belanger:2014vza} for relic density, in the following way:

\begin{enumerate}

\item We first compute the DM relic density as a function of tree-level DM mass $m_{\rm{DM}}$ keeping tree-level Higgs mass $m_h$ at its experimentally measured value~\cite{ParticleDataGroup:2024cfk}, see plots for ``Tree-Level" computations. For each value of DM mass, we compute the Higgs-DM coupling strength $\lambda_{H \rm{DM}}$, for example, see Fig.~\ref{fig:SSDM_RDtree}. 

\item We use these values to compute, within $\overline{\rm{MS}}$ scheme, the two-loop corrections to the Higgs mass $\Delta^2_h = M^2_h - m^2_h$ and one-loop corrections to DM mass $\Delta^2_{\rm{DM}} = M^2_{\rm{DM}} - m^2_{\rm{DM}}$ as well as RGE for all SM and BSM parameters. The procedure for these computations is already explained in the previous section. Here, $m_i$ and  $M_i$; $i = \{h, \rm{DM}\}$ denote the tree-level and loop corrected masses, respectively.

\item Since the Higgs self-quartic coupling $\lambda_H$ is not yet measured, we use this freedom to adjust its tree-level value such that the two-loop corrected Higgs mass $M_h$ is always within its experimentally measured value. This means that the tree-level value $m_h$ is changed appropriately with changing DM mass. 

\item We use these values to again compute DM relic density with changing loop corrected DM mass $M_{\rm{DM}}$  but now using loop corrected masses and RGE evolved couplings, see ``Loop-Level" plots. 

\item We stop when we reach the conformal limit of $m_h \to 0$ where the loop contribution equals the experimentally observed Higgs mass\footnote{Note that an alternative natural stopping point is $\Delta^2_h = m^2_h = M^2_h/2 = 88.53^2 \, \, \rm{GeV^2}$ as beyond this point, the loop correction becomes larger than the tree-level value, signaling a breakdown of the perturbative computation. This scenario yields similar albeit bit more stringent limits on the viable range for the DM mass and does not change our main conclusions.}.

\end{enumerate}

In this section, we implement the above mentioned procedure for three simple DM models that involve an additional scalar field, namely, the singlet scalar models--real and complex, and the inert doublet model. Similar constraints are also applicable to a wide variety of DM models which we intend to examine case by case in follow up works.

\subsection{Real singlet scalar DM}
\label{Subsec:relic:RSM}

One of the simplest extensions of the SM to accommodate the DM candidate is the singlet scalar DM model. In this framework, the SM is extended by adding the DM candidate $S$ which is a real scalar and transforms trivially under the SM gauge group. For the stability of DM, the model has a $Z_2$ symmetry with the following charge assignment: $Z_2 \ (S)=-1$, $Z_2$ (SM) = 1. The scalar $S$ does not acquire any VEV to ensure that $Z_2$ remains unbroken. The scalar potential for the real singlet scalar model is given as: 
\begin{equation}
    V_{\rm RSM}=\,\mu_H^2\, H^\dagger H + \frac{\lambda_{H}}{2}\,(H^\dagger H)^2+\frac{1}{2}\,\mu_S^2\, S^2 + \frac{\lambda_{HS}}{2}\,S^2\,H^\dagger H+\frac{\lambda_S}{2}\, S^4\,.
    \label{Eq:RSM}
\end{equation}
Where $H$ is the SM Higgs doublet and $\lambda_i$ are the quartic couplings. Note that the coupling between the Higgs and DM, $\lambda_{HDM} \equiv \lambda_{HS}$, is the only interaction responsible for DM annihilation. Additionally, its strength also determines the magnitude of the DM-induced loop correction to the Higgs mass.

\begin{center}
    \uline{\bf {Tree - Level}} 
\end{center}
We start with the tree-level computation for relic density. Let's first see the theoretical constraints and mass spectrum of scalars in this model. 
The requirement to have a stable minimum for the potential in Eq.~\ref{Eq:RSM} at the tree-level implies the following conditions on the quartic couplings
\begin{align}
\lambda_H,\,\lambda_S\,>\,0,\quad\lambda_{HS}\,\ge\,-2\sqrt{\lambda_H\lambda_S} \ .
\end{align}
The ``naive" requirements of perturbativity, on the other hand, impose the following condition on the quartic couplings\footnote{As mentioned before the more accurate S-matrix computation of perturbativity typically allows for a more relaxed upper limit on Higgs-DM coupling, see \cite{Goodsell_2018}. Here, we restrict ourselves to the more conservative ``naive perturbative" limit. } 
\begin{eqnarray}
    |\lambda_i|\,\le\,4\pi \ .
\end{eqnarray}
After the electroweak symmetry breaking, SM Higgs doublet can be expanded as follows
\begin{eqnarray}
  H = \frac{1}{\sqrt{2}}\begin{pmatrix}
         \sqrt{2}G^{+}\\
        v + h + i G^{0} 
     \end{pmatrix} \ .
\end{eqnarray}     
where $G^+$ and $G^0$ are Goldstone bosons to be `eaten' by the $W^+$ and $Z$ bosons, respectively. The tree-level masses for the Higgs and the scalar $S$ take the following form:
\begin{eqnarray}
m_h^2 &=& \lambda_H v^2\, , \qquad
    m_S^2 \, = \, \mu_S^2+\frac{1}{2}\lambda_{HS}v^2 \ .
\end{eqnarray}

\begin{figure}[!h]
    \centering
    \includegraphics[width=0.45\linewidth]{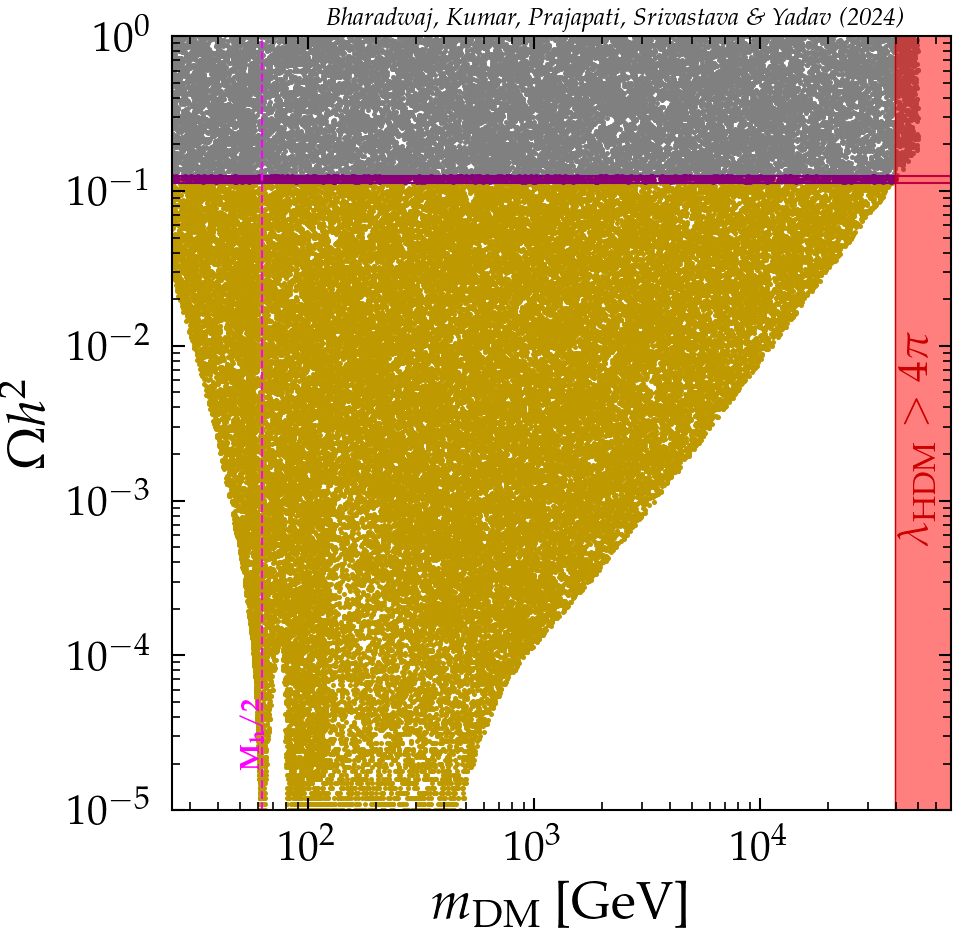}~~~
    \includegraphics[width=0.45\linewidth]{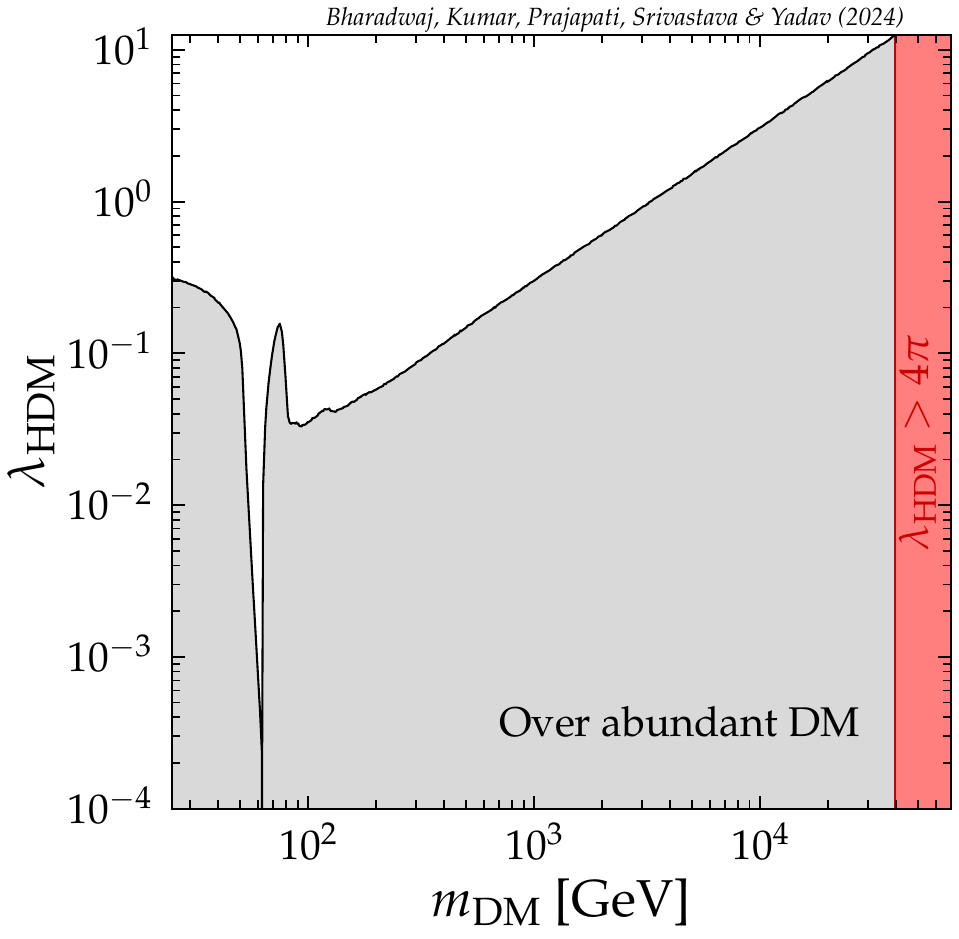}
    \caption{DM relic density (left panel) and the Higgs-DM coupling $\lambda_{\rm HDM}\equiv\lambda_{HS}$ (right panel) as a function of DM mass. Yellow (gray) points represent under (over) abundant DM relic density. The purple points satisfy the correct relic density \cite{Planck:2018vyg}. The red-shaded region is constrained by naive perturbativity limit.  
    }
    \label{fig:SSDM_RDtree}
\end{figure}

In the left panel of Fig.~\ref{fig:SSDM_RDtree}, we show the relic density as a function of DM mass.
Here, the yellow (gray) points represent the under (over) abundant DM relic density. The purple points satisfy the correct DM relic abundance in accordance with the Planck measurement~\cite{Planck:2018vyg}.  Notice that in this simple model, the DM couples to SM particles only via Higgs.  Thus, the DM relic abundance is solely controlled by the $\lambda_{HS}$ coupling. The requirement of correct relic density determines $\lambda_{HS}$ for a particular tree-level DM mass, $m_S \equiv m_{\rm{DM}}$, as shown in the right panel of Fig.~\ref{fig:SSDM_RDtree}. 
As the DM mass increases, $\lambda_{HS}$ also needs to be increased to satisfy the correct relic density. Our analysis shows that when $m_S$ reaches approximately 40 TeV, the required $\lambda_{HS}$ approaches the perturbativity limit, with $\lambda_{HS} \approx 4 \pi$.
 Therefore, the naive perturbativity requirement for $\lambda_{HS}$ constraints the region with  DM heavier than approximately 40 TeV. Note that, other theoretical constraints, such as the unitarity bound, are less stringent than perturbativity and permit DM with masses up to about 100 TeV~\cite{Griest:1989wd,Smirnov:2019ngs}. 

\begin{figure}[!h]
    \centering
    \includegraphics[width=0.45\linewidth]{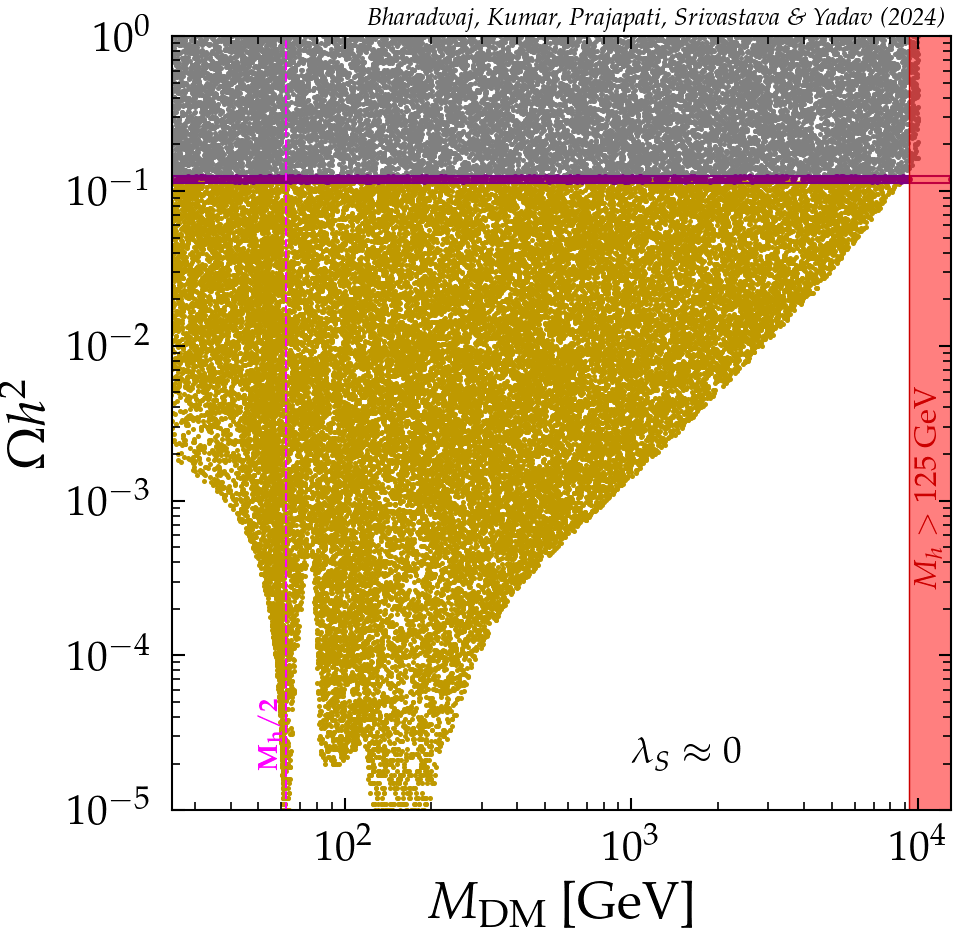}~~~  
    \includegraphics[width=0.45\linewidth]{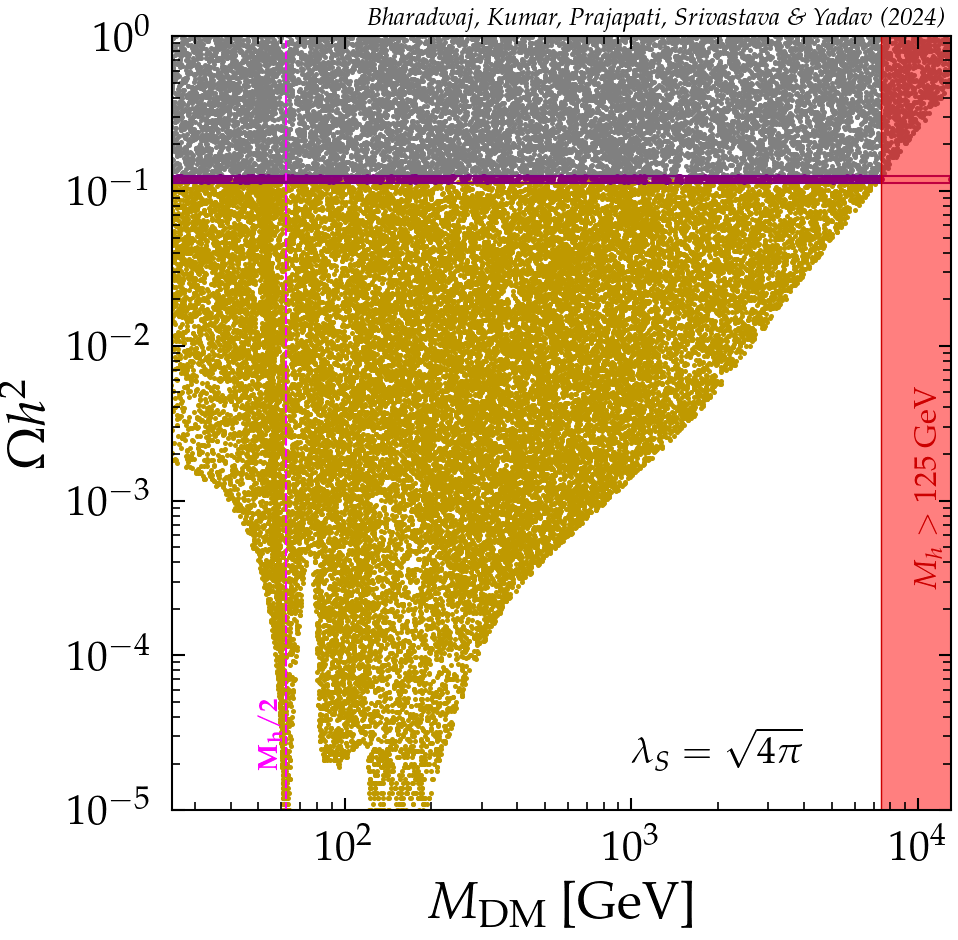}
    \caption{ DM relic density as a function of loop corrected DM mass for $\lambda_s \approx 0$ (left) and $\lambda_s \approx \sqrt{4 \pi}$ (right).  The red-shaded region does not satisfy the experimentally observed value of Higgs mass. The color code for the scatter points is the same as Fig.~\ref{fig:SSDM_RDtree}.}
    \label{fig:SSDM_RD}
\end{figure}

\begin{center}
    \uline{\bf {Loop - Level}}
\end{center}
%
 When the loop corrections to the Higgs mass are taken into account, the picture changes significantly. Since the loop corrections to the Higgs mass are directly proportional to $\lambda_{HS}$ and $m_S$, these corrections grow rapidly with increasing DM mass. In the high-$m_S$ regime, the loop corrections to the Higgs mass become comparable to its observed value.
Consequently, the measured Higgs mass, $M_H = 125.20\pm0.11$ GeV~\cite{ParticleDataGroup:2024cfk}, imposes stringent constraints on the parameter space. Specifically, regions where the loop corrections exceed the observed value of Higgs mass are excluded by this experimental measurement.
 This exclusion is illustrated in Fig.~\ref{fig:SSDM_RD}, which shows the relic density as a function of the loop corrected DM mass, $M_{DM} \equiv M_S$. The vertical red-shaded region indicates the parameter space ruled out due to excessive two-loop corrections to the Higgs mass. As we can see, the requirement of the measured Higgs mass provides much stronger constraints than the naive perturbative criterion. In this scenario, the maximum allowed parameter space reduces from roughly 40 TeV to nearly 9.3 TeV. 

An important aspect of this analysis is the role of the DM self-quartic coupling, $\lambda_S$. As $\lambda_S$ contributes to the Higgs mass corrections at the two-loop-level, larger values of $\lambda_S$ result in more significant corrections to the Higgs mass. Consequently, for substantial $\lambda_S$, the loop corrections to Higgs mass exceed the observed value of Higgs mass at a slightly lower DM mass compared to scenarios where $\lambda_S$ is negligible. This effect is evident from the shift 
in the vertical red-shaded region between the two panels of Fig.~\ref{fig:SSDM_RD}. When $\lambda_S$ is set to nearly zero, a DM mass above 9.3 TeV (Fig.~\ref{fig:SSDM_RD}, left panel) fails to reproduce the observed Higgs mass for correct relic points. On the other hand, if $\lambda_S = \sqrt{4\pi}$, a DM mass exceeding 7.4 TeV (Fig.~\ref{fig:SSDM_RD}, right panel) is unable to reproduce the observed Higgs mass for the correct relic points.
 
\FloatBarrier

\subsection{Complex singlet scalar DM}
\label{Subsec:relic:CSM}

Another simple extension of SM to accommodate the DM candidate is the complex singlet scalar model where the SM is extended with a complex scalar, $\zeta$, which transforms trivially under the SM gauge group. It is also charged under a global $U(1)$ symmetry, which ensures its stability. The scalar potential for the complex singlet scalar model is given by: 
\begin{equation}
    V_{\rm CSM}=\,\mu_H^2\, H^\dagger H + \frac{\lambda_{H}}{2}\,(H^\dagger H)^2\,+\,\mu_\zeta^2\,\zeta^*\zeta + \lambda_{H\zeta}\,\zeta^*\zeta H^\dagger H+\frac{\lambda_\zeta}{4} (\zeta^*\zeta)^2 \ .
    \label{Eq:CSM}
\end{equation}
Here, as before, $H$ is the SM Higgs doublet, and $\lambda_i$ are the quartic couplings. The Higgs-DM coupling, $\lambda_{H\zeta}$ controls the DM relic density and also contributes to the loop corrections to the Higgs mass.
\begin{center}
    \uline{\bf {Tree - Level}} 
\end{center}
%
\begin{figure}[!h]
    \centering
    \includegraphics[width=0.45\linewidth]{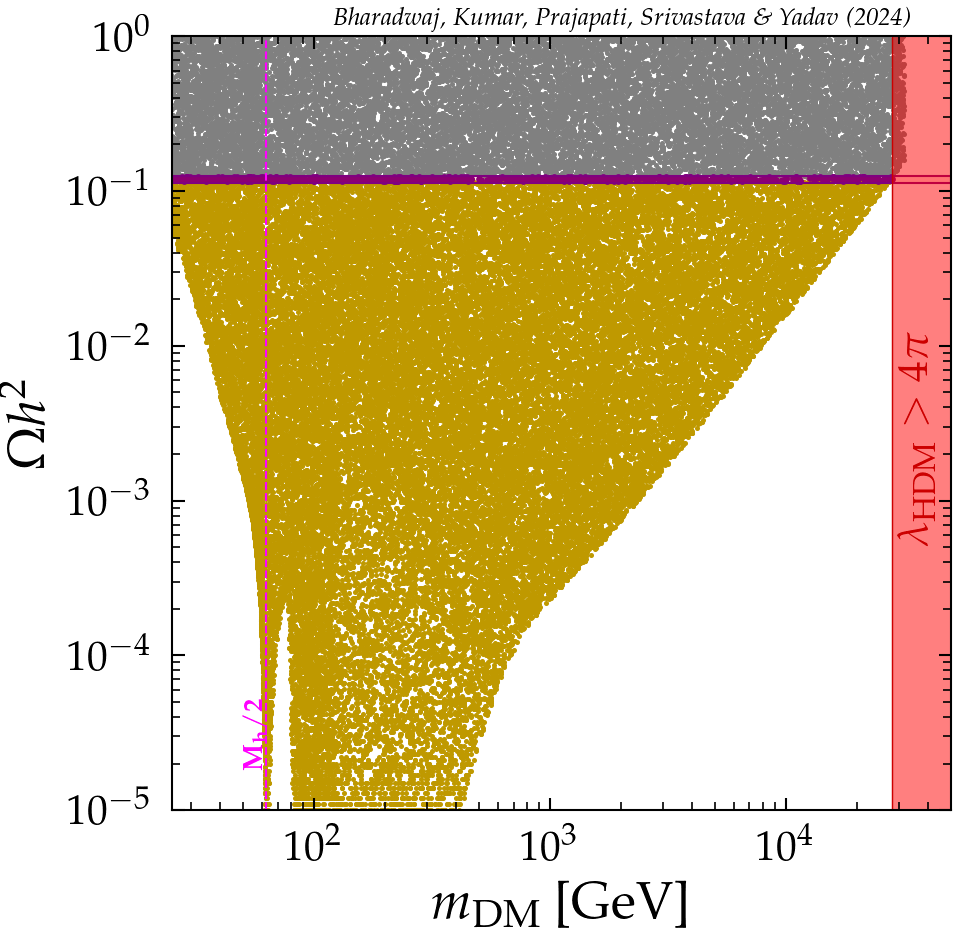}~~~
    \caption{DM relic density as a function of DM mass. The color code is the same as Fig.~\ref{fig:SSDM_RDtree}. The red-shaded region does not satisfy the perturbativity constraints.}
    \label{fig:CSM_RDT}
\end{figure}
The requirement to have a stable minimum for the potential in Eq.~\ref{Eq:CSM} at the tree-level implies the following conditions on the quartic couplings
\begin{align}
\lambda_H,\,\lambda_\zeta\,>\,0,\quad\lambda_{H\zeta}\,\ge\,-\sqrt{\lambda_H\lambda_\zeta/2} \ .
\end{align}
The naive perturbativity requirements can be fulfilled by demanding the following conditions on the quartic couplings 
\begin{eqnarray}
    |\lambda_i|\,\le\,4\pi \ .
\end{eqnarray}
After the electroweak symmetry breaking, the tree-level mass of the Higgs and the complex scalar can be written as
\begin{eqnarray}
 m_h^2 &=& \lambda_H v^2\, , \qquad   m_\zeta^2 \, = \, \mu_\zeta^2+\frac{1}{2}\lambda_{H\zeta}\,v^2 \ .
\end{eqnarray}

Following an analysis similar to the real singlet case, we find that when the Higgs mass is computed at tree-level, the Higgs-DM coupling $\lambda_{H\zeta}$ becomes non-perturbative for DM with mass exceeding approximately 30 TeV, as shown in Fig. \ref{fig:CSM_RDT}. Consequently, the DM mass in the complex singlet case cannot exceed $\sim$ 30 TeV.

\begin{center}
    \uline{\bf {Loop - Level}} 
\end{center}
 Similar to the real singlet case, when the Higgs mass is computed at two-loop-level, we observe that the loop corrections to the Higgs mass become comparable to the observed value of Higgs mass around a DM mass of 5.1 TeV. Consequently, the parameter space beyond this point is excluded as it fails to respect the observed value of the Higgs mass. Therefore, the constraints on the Higgs mass impose significantly stricter bounds than those obtained from naive perturbativity limits, reducing the upper limit of the viable parameter space from around 30 TeV to approximately 5.1 TeV.
 \begin{figure}[!h]
    \centering
    \includegraphics[width=0.45\linewidth]{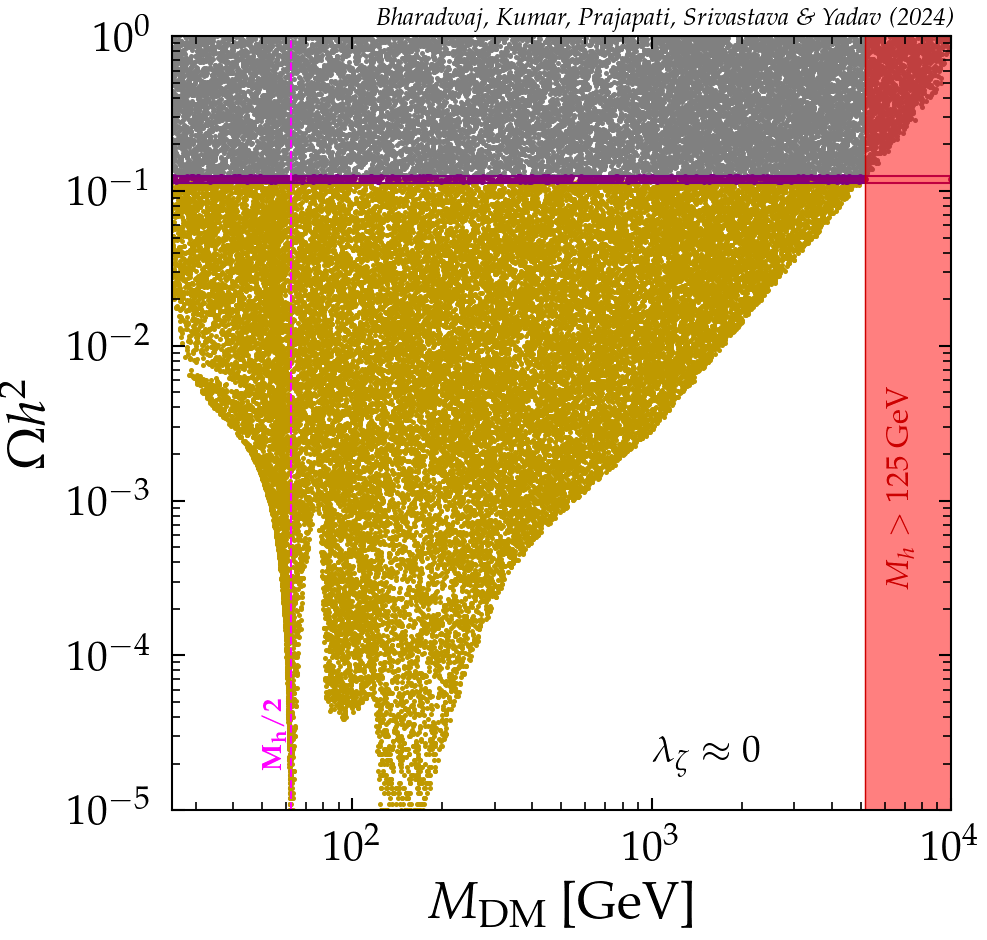}~~~    \includegraphics[width=0.45\linewidth]{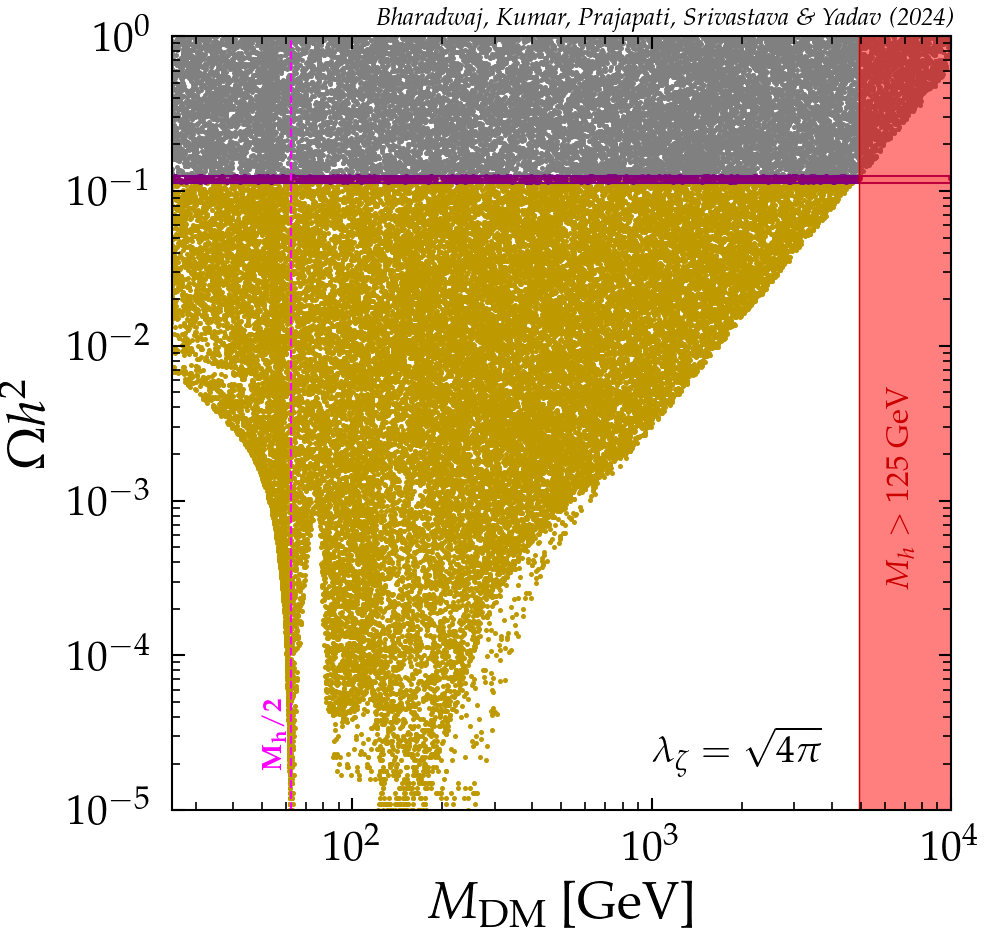} 
    \caption{DM relic density as a function of loop corrected DM mass for $\lambda_\zeta \approx 0$ (left) and $\lambda_\zeta \approx \sqrt{4 \pi}$ (right).  The red-shaded region does not satisfy the observed Higgs mass, computed at two-loop-level.}
    \label{fig:CSM_RD0}
\end{figure}
In Fig.~\ref{fig:CSM_RD0}, we present the relic density vs DM mass plot. Again we plot it for two different values of DM self-quartic coupling, $\lambda_{\zeta}$.
 When $\lambda_\zeta$ is set to approximately zero, a DM mass above 5.1 TeV (Fig.~\ref{fig:CSM_RD0}, left panel) fails to reproduce the observed Higgs mass for correct relic points. Similarly, if $\lambda_\zeta = \sqrt{4\pi}$, a DM mass exceeding 4.9 TeV (Fig.~\ref{fig:CSM_RD0}, right panel) is unable to reproduce the observed Higgs mass for correct relic points.

\FloatBarrier

As aforementioned, in both the real and complex singlet DM models, the Higgs-DM quartic coupling solely controls the DM annihilation into SM particles.
However, there are other models where DM couples to additional SM particles, such as the W and Z bosons, and DM (co-)annihilation through W,Z channels can also contribute to the DM relic density. One such model is the inert doublet model, where DM is the neutral component of an $SU(2)_{L}$ doublet and couples to both the W and Z bosons. We discuss its relic density aspects below.

\subsection{Inert Doublet Scalar DM}
\label{Subsec:relic:IDM}

In the inert doublet model, an additional scalar $SU(2)_L$ doublet, $\eta$, is introduced alongside the SM fields. The neutral component of this scalar doublet, $\eta$, serves as a DM candidate in the model. To ensure the stability of the DM, a global ${Z}_2$ symmetry is imposed, under which $\eta$ is odd, while all SM fields are even. This symmetry prevents the decay of $\eta$, thereby stabilizing the DM candidate. The scalar potential for the inert doublet model is given as: 

\begin{eqnarray}
    V_{\rm IDM} & = & \mu_H^2 H^\dagger H + \mu_{\eta}^2 \eta^\dagger \eta + \frac{\lambda_{1}}{2} (H^\dagger H)^2 + \frac{\lambda_{2}}{2} (\eta^\dagger \eta)^2 + \lambda_{3} (H^\dagger H)(\eta^\dagger \eta) + \lambda_{4} (H^\dagger \eta)(\eta^\dagger H) \nonumber \\
    & + & \left[\frac{\lambda_{5}}{2} (H^\dagger \eta)^2 + h.c. \right].
    \label{Eq:IDM}
\end{eqnarray}
Here, $\mu_H^2$, $\mu_{\eta}^2$ and $\lambda_i$, $i \in \{1, 2,3,4,5\}$ are real parameters. Note that $\lambda_1$ is the Higgs self-quartic coupling whereas $\lambda_2$ is the self-quartic coupling of the $\eta$ doublet. The couplings $\lambda_3, \lambda_4, \lambda_5$ couple DM to the Higgs.

\begin{center}
    \uline{\bf {Tree - Level}} 
\end{center}
The requirement to have a stable minimum for the potential at the tree-level implies the following conditions on the scalar potential parameters \cite{Kannike:2016fmd}
\begin{eqnarray}
\lambda_{1}, \lambda_{2}  >  0, \hspace{0.5cm} \lambda_{3} > -2\sqrt{\lambda_1\lambda_2} \ , \hspace{0.6cm} \lambda_{3} + \lambda_4 - |\lambda_5| > -2\sqrt{\lambda_1\lambda_2}\ .
    \label{eq:vacuumstability}
\end{eqnarray}
While the naive perturbativity of the quartic couplings implies

\begin{equation}
   |\lambda_i| \leq {4 \pi} \ .
    \label{eq:perturbativity1}
\end{equation}
After the electroweak symmetry breaking, SM Higgs doublet and $\eta$ can be expanded as follows
\begin{eqnarray}
& & \begin{aligned}
  H = \frac{1}{\sqrt{2}}\begin{pmatrix}
         \sqrt{2}G^{+}\\
        v + h + i G^{0}
     \end{pmatrix}, & \hspace{1cm}& \eta = \frac{1}{\sqrt{2}}\begin{pmatrix}
\sqrt{2}\eta^+\\
\eta_{R}+i\eta_{I}
\end{pmatrix} \ .
\end{aligned}
\end{eqnarray}
Notice that the doublet $\eta$ does not get any VEV, hence the model is called the inert doublet model.
We can now compute the tree-level masses of the physical scalar states after symmetry breaking.
\begin{equation}
    m_{h}^{2} = \lambda_{1}v^2,
\quad
    m_{\eta^{\pm}}^{2} = \mu_{\eta}^{2} + \frac{\lambda_{3}}{2}v^2,
    \label{eq:metaplus}
\end{equation}
\begin{equation}
    m^2_{\eta_{I}} = \mu_{\eta}^{2} + (\lambda_{3}+\lambda_{4}-\lambda_{5})\frac{v^2}{2},
    \label{eq:metaplus}
\end{equation}
\begin{equation}
    m^2_{\eta_{R}} = \mu_{\eta}^{2} + (\lambda_{3}+\lambda_{4}+\lambda_{5})\frac{v^2}{2} \equiv \mu_{\eta}^{2} + \lambda_{345}\frac{v^2}{2} \ .
    \label{eq:metaplus}
\end{equation}
We consider the real part of the neutral scalar component as our DM candidate which requires that the masses of the $\eta$ components should be $m_{\eta_R} < m_{\eta_I}, m_{\eta^{\pm}}$\footnote{Alternatively, the imaginary part $\eta_I$ can also be taken as DM candidate provided it is the lightest component of the $\eta$ doublet. Our results will remain practically unchanged even for this choice.}. This requirement imposes the following conditions on scalar potential couplings.
\begin{equation}
    \lambda_4 + \lambda_5 \, < \, 0, \hspace{0.5cm} \lambda_5 \, < \,0.
    \label{etrDM}
\end{equation}

\begin{figure}[!h]
 \centering
        \includegraphics[height=8.4cm]{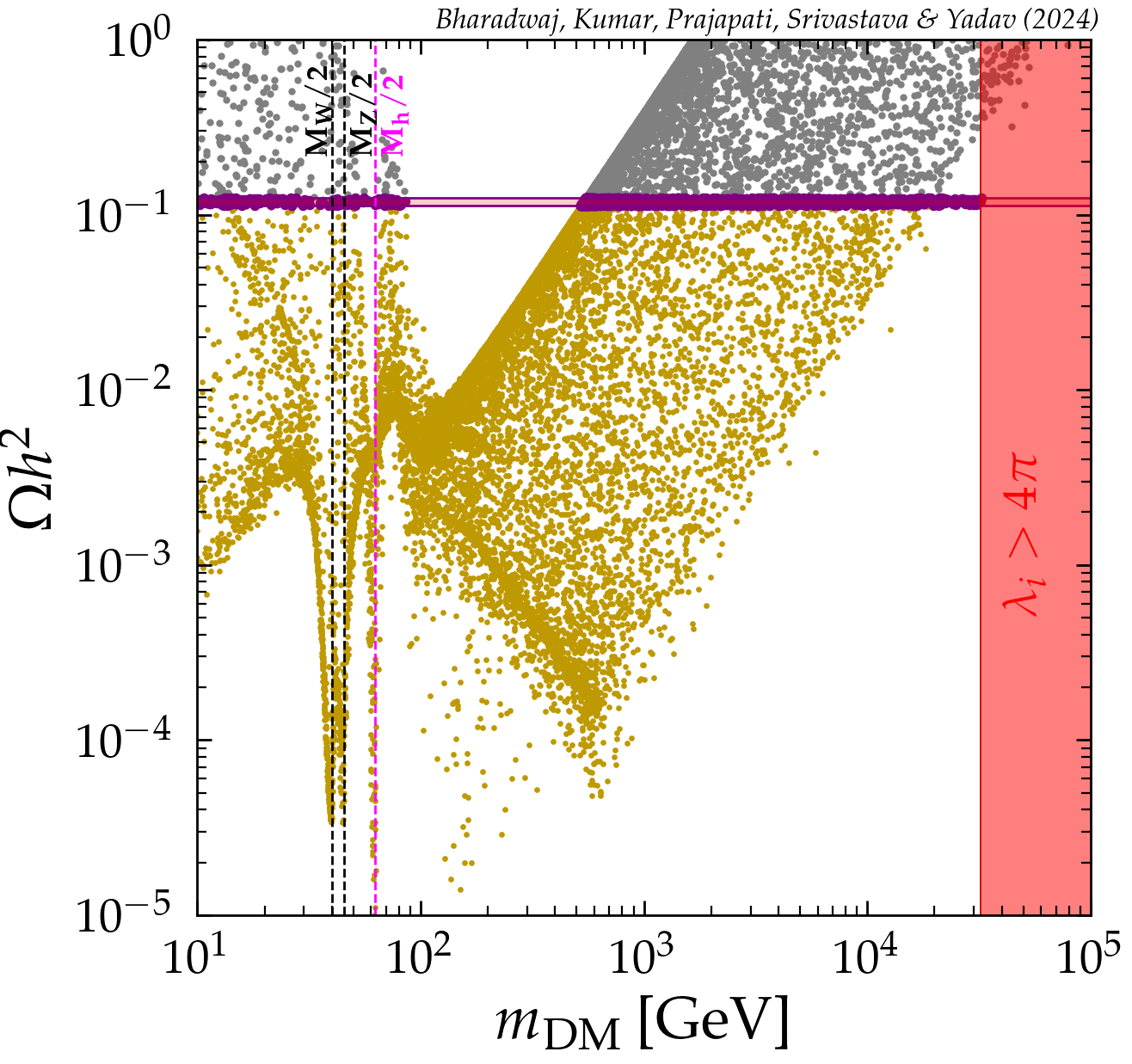}
        \caption{Relic density vs mass of neutral doublet scalar where Higgs mass is computed at tree-level. The color code is the same as Fig.~\ref{fig:SSDM_RDtree}. } 
        \label{fig:InertDM_Without_Loop}
\end{figure}

Fig.~\ref{fig:InertDM_Without_Loop} shows the dependence of the DM relic abundance on the DM mass $(m_{\rm{DM}}\equiv  m_{\eta_{R}})$.
The yellow (gray) points in the plot indicate regions where the DM $\eta_R$ is under (over) abundant. The purple points correspond to regions that satisfy the correct relic density \cite{Planck:2018vyg}. The red-shaded region in Fig.~\ref{fig:InertDM_Without_Loop} represents the parameter space where at least one of the couplings $\lambda_i$; $i = 1,2,3,4,5$ exceeds $4\pi$ for the points that satisfy the correct relic density. Compared to the previous two cases, the inert doublet model involves three Higgs-DM couplings, $\lambda_3$, $\lambda_4$, and $\lambda_5$, which significantly contribute to the computation of the relic density, with their combination $\lambda_{345} = \lambda_3 + \lambda_4 + \lambda_5$ playing a crucial role. We obtain two distinct mass regions that satisfy the correct relic density: The first regime from 10 GeV to 80 GeV\footnote{A singnificant part of this low mass parameter space gets ruled out by other constraints as dicussed in Sec.~\ref{Subsec:dd:IDM}.} and the second regime from 500 GeV to approximately 32 TeV. 
In contrast to the scalar DM-real and complex models, here, in the intermediate mass region, none of the points satisfies the correct relic abundance. This is due to the efficient annihilation mediated by the exchange of Higgs, W, and Z bosons.

When considering the particle masses at the tree-level, we find that  DM with a mass up to approximately 32 TeV is permissible. Beyond this mass, $\lambda_i$ couplings become non-perturbative in order to satisfy the observed relic density. Therefore, the naive perturbativity requirement for scalar potential couplings excludes the region heavier than approximately 32 TeV for inert doublet DM case.

\FloatBarrier

\begin{figure}[th]
 \centering
        \includegraphics[height=7.5cm]{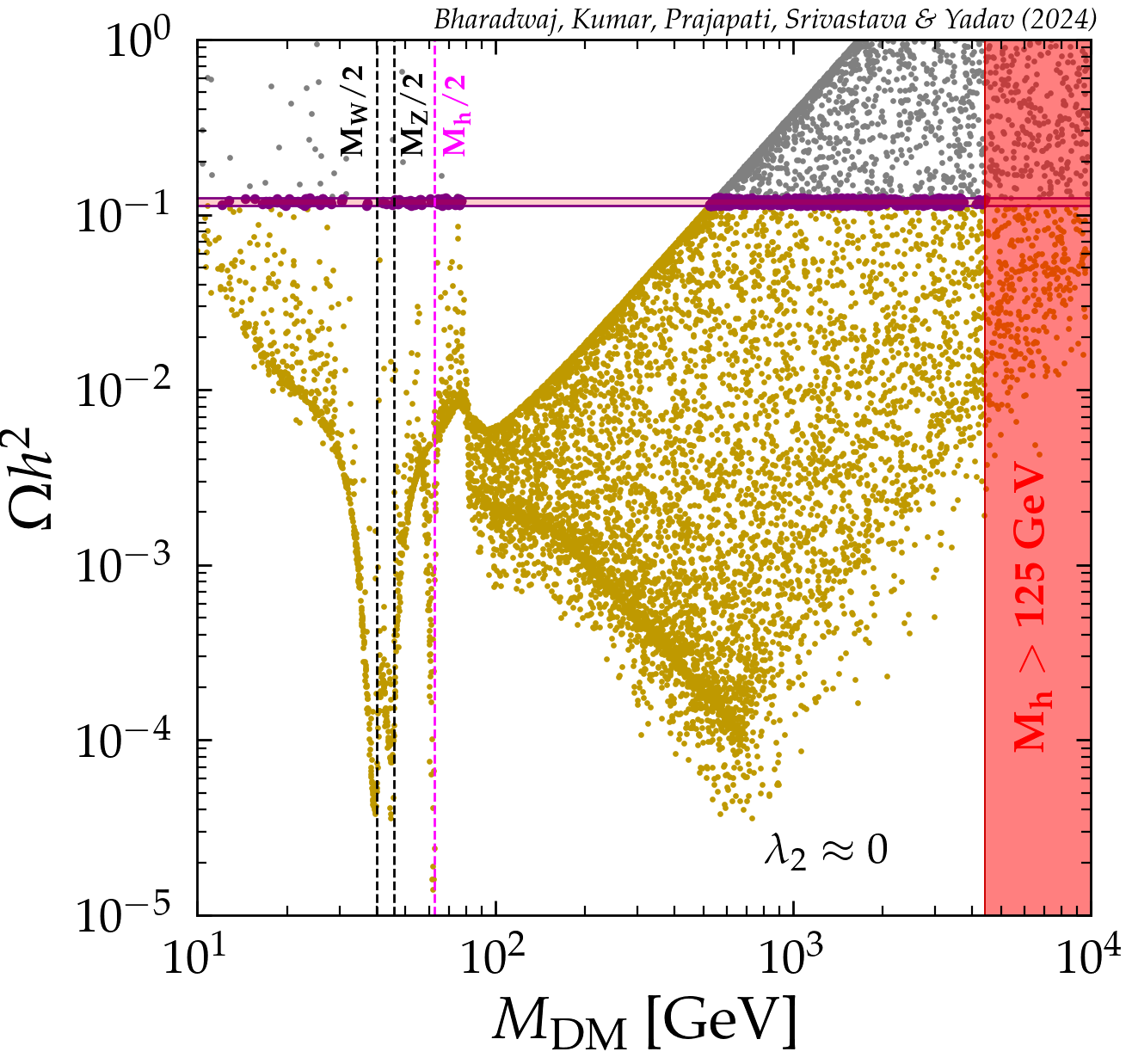}
        \includegraphics[height=7.5cm]{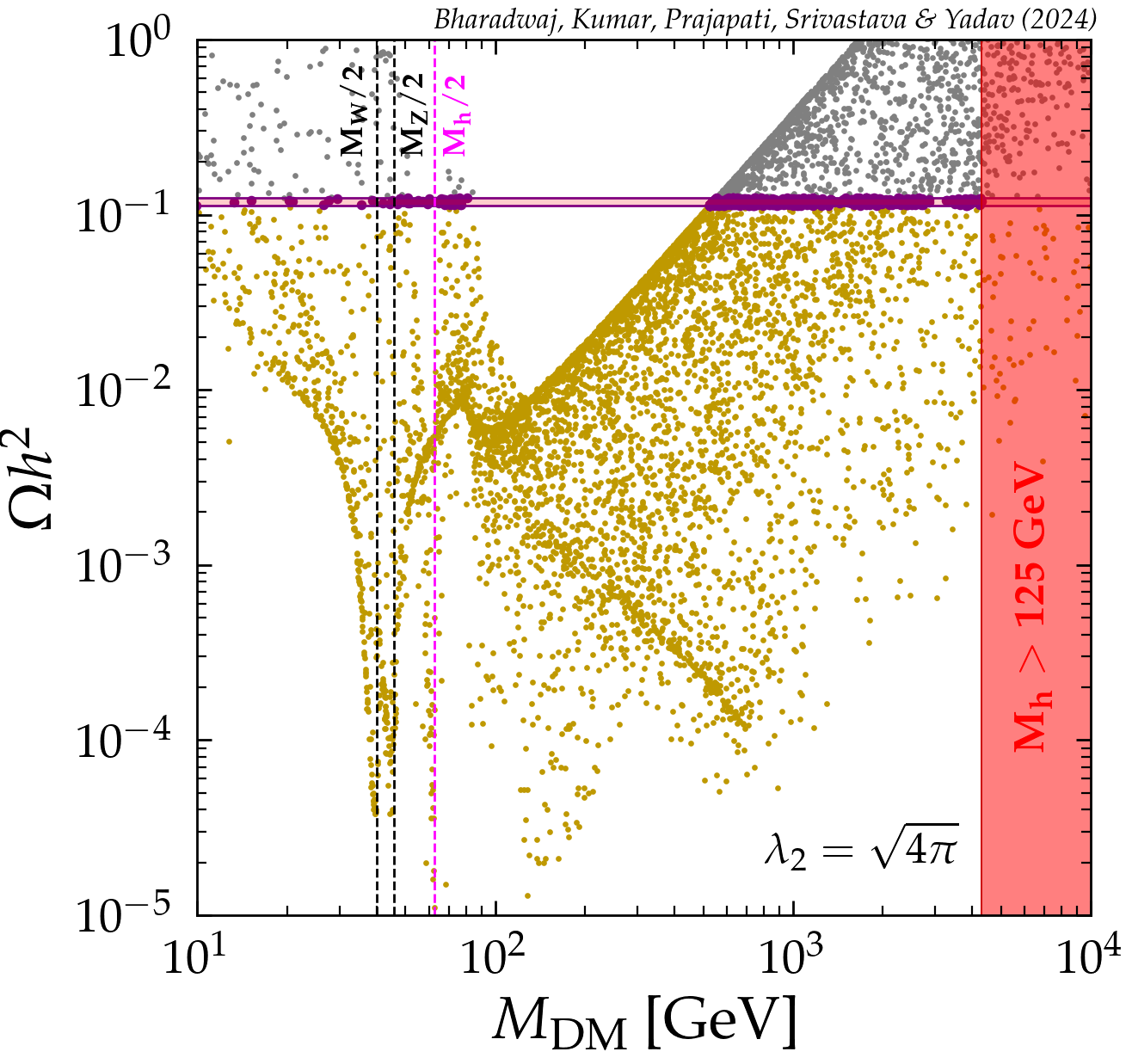}
        \caption{Relic density vs mass of neutral doublet scalar where Higgs mass is computed at two-loop-level with $\lambda_2$ $\approx$ 0 (left) and $\lambda_2$ = $\sqrt{4\pi}$ (right). The red-shaded region does not satisfy the measured value of Higgs mass.}
        \label{fig:InertDM_With_Loop}
\end{figure}

\begin{center}
    \uline{\bf {Loop - Level}} 
\end{center}
%
Again, this parameter space changes significantly when Higgs two-loop corrections are taken into account. In
Fig.~\ref{fig:InertDM_With_Loop}, we show the DM relic density as a function of the DM mass $(M_{\rm{DM}}\equiv  M_{\eta_{R}})$, when the Higgs mass is computed at the two-loop-level. To examine the impact of the DM self-quartic coupling, $\lambda_2$ is set to approximately zero in the left panel, while in the right panel, $\lambda_{2}=\sqrt{4\pi}$ has been taken. 
%
We find that the loop corrections to the Higgs mass become comparable to the observed value of Higgs mass around the DM mass of approximately 4.4 TeV for correct relic points. Beyond this threshold, these loop corrections become too large to be compensated by tuning the Higgs self-quartic coupling. Consequently, the parameter space where correct relic density is satisfied for DM masses exceeding 4.4 TeV fails to reproduce the observed Higgs mass. Our analysis indicates that the effect of the DM self-quartic coupling 
$\lambda_2$ on the results is minimal. When $\lambda_2$ is set to approximately zero, a DM mass above 4.4 TeV (Fig.~\ref{fig:InertDM_With_Loop}, left panel) fails to reproduce the observed Higgs mass for correct relic points. Similarly, if 
$\lambda_{2}=\sqrt{4\pi}$, a DM mass exceeding 4.3 TeV (Fig.~\ref{fig:InertDM_With_Loop}, right panel) is unable to reproduce the observed Higgs mass for correct relic points. 
The combination of all relevant constraints leads to an allowed high mass range of  $M_{\rm{DM}} \sim (0.5-4.4)$ TeV for this scenario. Therefore, the constraints on the Higgs mass impose much stricter bounds compared to naive perturbativity limits, reducing the maximum allowed parameter space in this case from about 32 TeV to approximately 4.4 TeV.

In summary, as the DM mass increases, the correct relic density can only be obtained by a corresponding increase in the Higgs-DM coupling. This puts a lower limit on the allowed Higgs-DM interaction strength and implies that the larger the DM mass, the larger should be the Higgs-DM coupling. Now both DM mass and Higgs-DM couplings control the magnitude of DM-induced loop correction to Higgs mass. Thus, for a sufficiently large DM mass, the DM-induced correction to the Higgs mass becomes so significant that the Higgs mass exceeds its experimental value, even in the conformal limit ($m_h \to 0$), leading to an upper bound on the DM mass of around a few TeVs.

\FloatBarrier

\section{Direct Detection of WIMP-nucleon scattering} \label{sec:dd}

 In the previous section, we discussed that parameter space satisfying correct relic density for DM masses more than a few TeVs is ruled out by Higgs mass constraints. In this section, we look at its impact on the DM search in the direct detection experiments.
In order to obtain a complete picture, in our analysis we also include all the other applicable constraints on DM parameter space as follows:
\begin{enumerate}
\item We are imposing the constraints from the oblique parameters at the $1\sigma$ level, as provided by the PDG \cite{ParticleDataGroup:2024cfk}.
\begin{equation}
    S = -0.04 \pm 0.10, \,  T = 0.01 \pm 0.12,  \, U = -0.01 \pm 0.09.
\end{equation}
\item We are imposing the bound on the branching ratio of Higgs invisible decay  to the dark sector particles
i.e. $Br (h \to \rm{DM}\, \rm{DM}) < 0.107$, from the recent LHC data \cite{ATLAS:2023tkt}.
\item For the inert doublet DM case, we also impose the LEP constraints on the neutral and charged components of the doublet scalar, as detailed in \cite{Belyaev:2016lok,Cao:2007rm}.
\end{enumerate}

Let us now look at the implications of imposing all these constraints on the three models discussed in the previous section. 

\FloatBarrier

\subsection{Real Singlet Scalar DM}
\label{Subsec:dd:RSM}
%
For the real singlet scalar DM model described by Eq.~\eqref{eq:eftsdm}, the spin-independent scattering cross-section for WIMP-nucleon interactions is expressed as 
\begin{eqnarray}
 \sigma_{\rm scalar} & = & \frac{\lambda_{HS}^2 }{4\pi M^4_h}\,\frac{m^4_n}{(m_n + M_{\rm DM})^2} \, f^2_n \,.
\end{eqnarray}
Here, $m_n$ is the nucleon mass, $M_{\rm DM}$ is the loop corrected DM mass, $M_h$ is the loop corrected Higgs mass, and  $f_n$ is the hadronic matrix element for the Higgs-nucleon effective coupling.

\begin{figure}[h!]
\centering
\includegraphics[width=.49\textwidth]{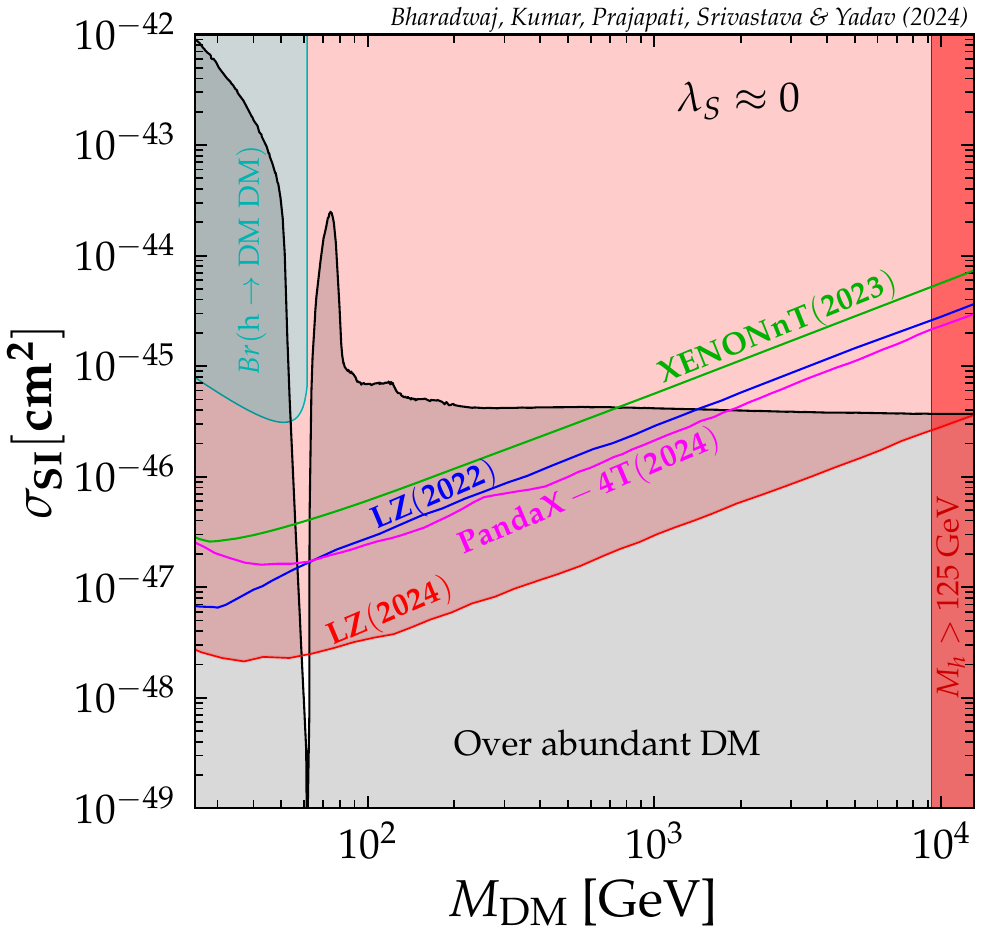}~~~
\includegraphics[width=.49\textwidth]{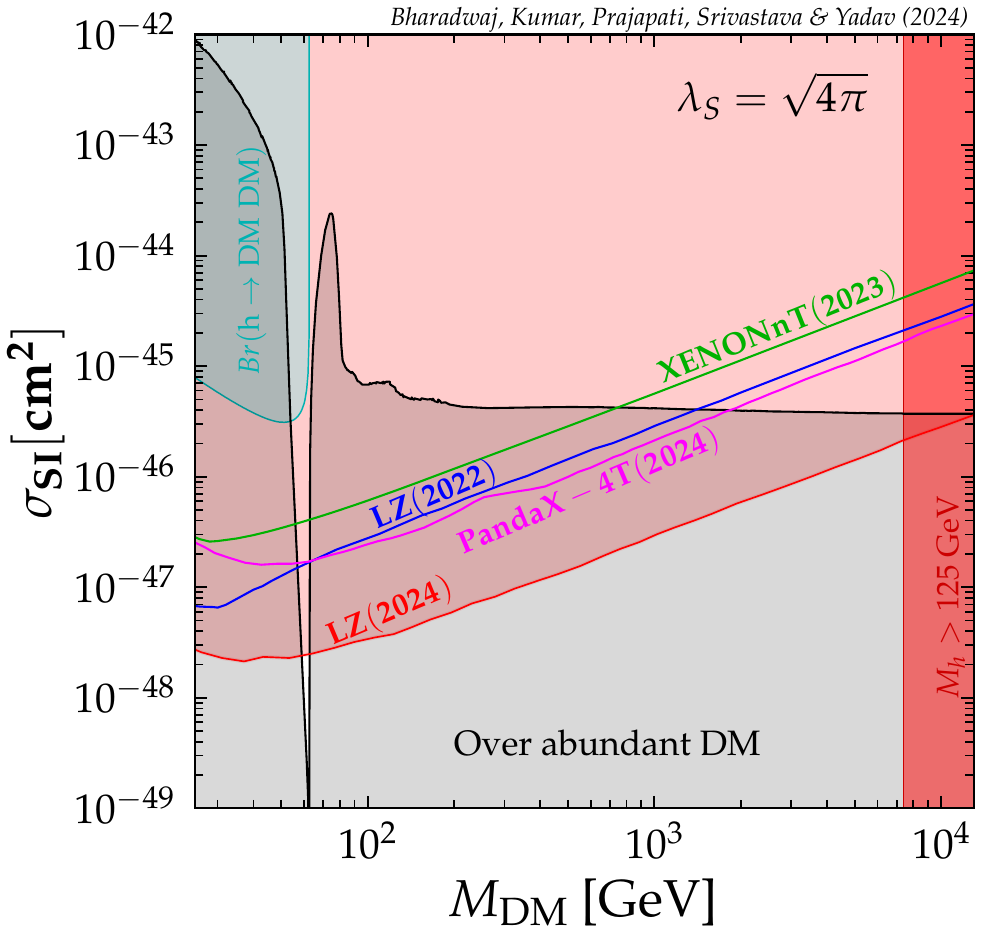}
\caption{Spin-independent nucleon-DM scattering cross-section as a function of the DM mass. Higgs mass is calculated at two-loop-level with $\lambda_S\approx0$ (left) and $\lambda_S=\sqrt{4\pi}$ (right). Various direct detection experimental limits are also shown by colored lines. The cyan region represents the constraint from the Higgs invisible decay~\cite{ATLAS:2023tkt}.}
\label{fig:RSM2}
\end{figure}

 In Fig.~\ref{fig:RSM2}, we present the spin-independent WIMP-nucleon scattering cross-section as a function of the DM mass, where Higgs mass is computed at the two-loop-level, with two different scenarios: $\lambda_S$ $\approx$ 0 in the left panel, and $\lambda_S$ = $\sqrt{4\pi}$ in the right panel. The gray-shaded region is excluded because the corresponding DM parameter space leads to over abundance. The solid black line represents the DM-nucleon scattering cross-section consistent with the correct relic abundance.
The cyan region corresponds to constraint from the invisible Higgs decay~\cite{ATLAS:2023tkt}, while the vertical red band indicates the parameter space excluded by loop-corrected Higgs mass constraints. As previously discussed, the Higgs mass correction reduces the allowed parameter space from 40 TeV to 9.3 TeV. Limits from various direct detection experiments, including XENONnT~\cite{XENON:2023cxc}, PandaX-4T~\cite{PandaX:2024qfu}, and LZ~\cite{LZ:2022lsv,LZCollaboration:2024lux}, are shown as distinct colored lines.  Additionally, notice that the recent LZ (2024) result has further reduced the allowed cross-section for a given DM mass by an order of magnitude. From Fig.~\ref{fig:RSM2}, it is evident that the combined effect of constraints from the Higgs mass correction and the recent LZ (2024) result rules out the entire DM parameter space, except for a narrow region near $ M_h/2 $.

\FloatBarrier

 \subsection{Complex Singlet Scalar DM}   \label{Subsec:dd:CSM}
%
For the complex singlet scalar model,
\begin{figure}[h!]
\centering
\includegraphics[width=.49\textwidth]{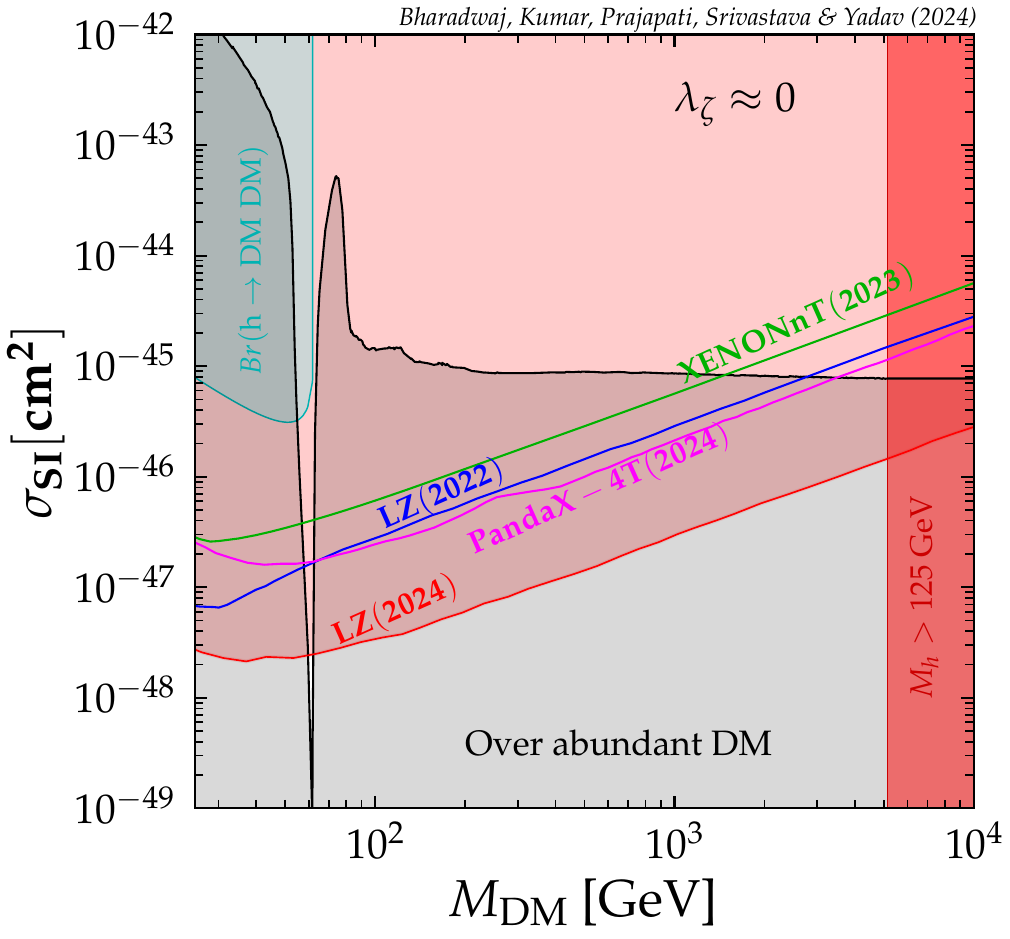}
\includegraphics[width=.49\textwidth]{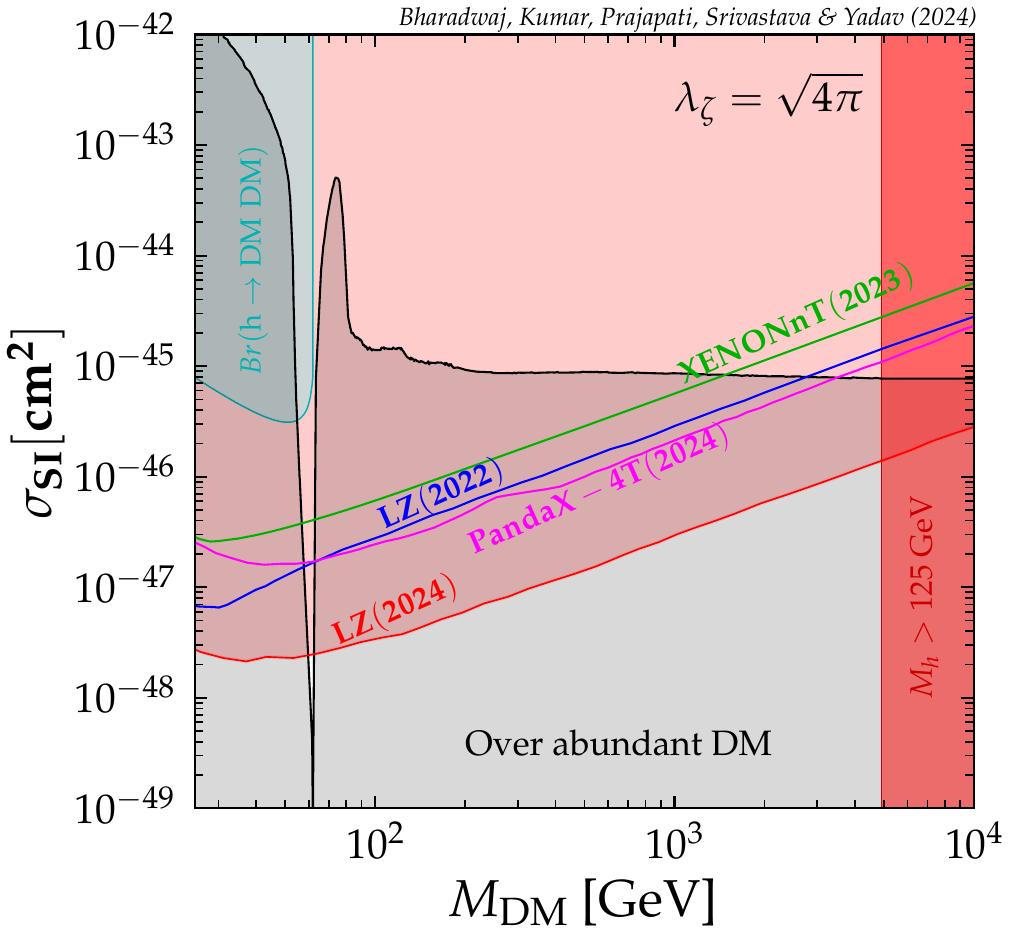}
\caption{Spin-independent nucleon-DM scattering cross-section as a function of DM mass. Higgs mass is calculated at two-loop-level with $\lambda_\zeta\approx0$ (left) and $\lambda_\zeta=\sqrt{4\pi}$ (right).}
\label{fig:CSM}
\end{figure}
 Fig.~\ref{fig:CSM} displays the spin-independent WIMP-nucleon scattering cross-section as a function of DM mass, where Higgs mass is computed at the two-loop-level.  Here, again we have plotted it for the two different values of the DM self-quartic coupling,
 $\lambda_\zeta$ $\approx$ 0 in the left panel, and $\lambda_\zeta$ = $\sqrt{4\pi}$ in the right panel. The color code remains the same as Fig.~\ref{fig:RSM2}. As before, in this case also,  the Higgs mass correction reduces the allowed parameter space from 30 TeV to 5.1 TeV. From Fig.~\ref{fig:CSM}, one can infer that the combined effect of constraints from the Higgs mass correction and the recent LZ (2024) result rules out the entire DM parameter space, except for a narrow region near $ M_h/2 $.

\FloatBarrier
 In the above two cases, we observed that the constraints from the Higgs mass corrections and the recent LZ (2024) result effectively exclude almost the entire parameter space. This conclusion, in general, should hold for various other models where DM annihilation through Higgs is the dominant source of achieving the observed relic abundance, as the same coupling governs both annihilation and direct detection processes of the DM candidates. However, in scenarios where co-annihilation channels are also present along with annihilation processes, the mass regime from around 500 GeV till few TeVs may remain viable.
These new channels may relax the constraints on the coupling responsible for the DM direct detection and hence, the parameter space in this mass-regime can be still allowed in such types of models. 
Note that even in these cases limits from DM-induced loop corrections to the Higgs mass are still applicable and still rule out DM masses above a few TeVs.
Let's discuss one such model, the inert doublet model, and its viable DM parameter space.
\subsection{Inert Doublet Scalar DM}
\label{Subsec:dd:IDM}

\begin{figure}[th]
 \centering
        \includegraphics[height=7.3cm]{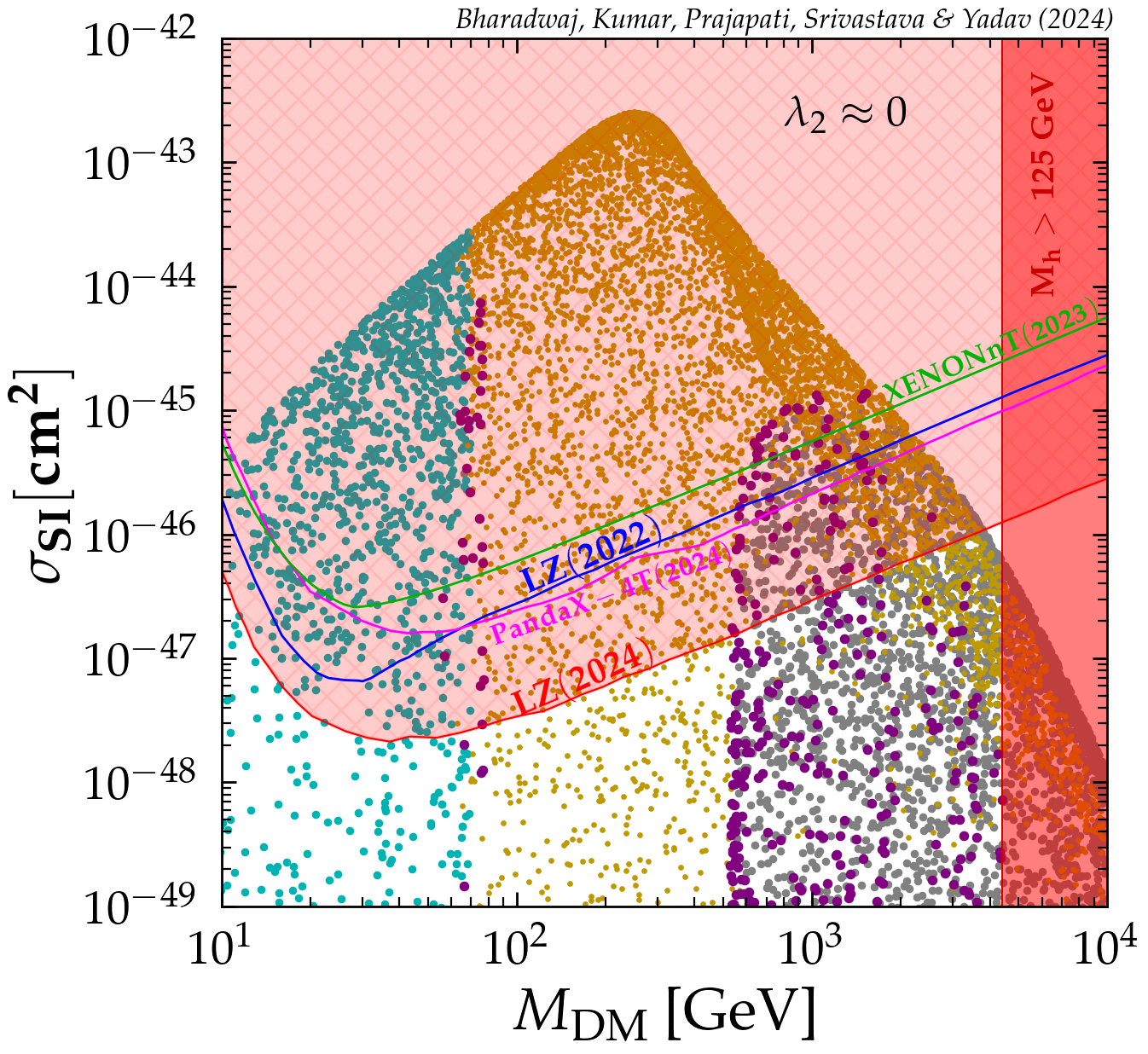}
        \includegraphics[height=7.3cm]{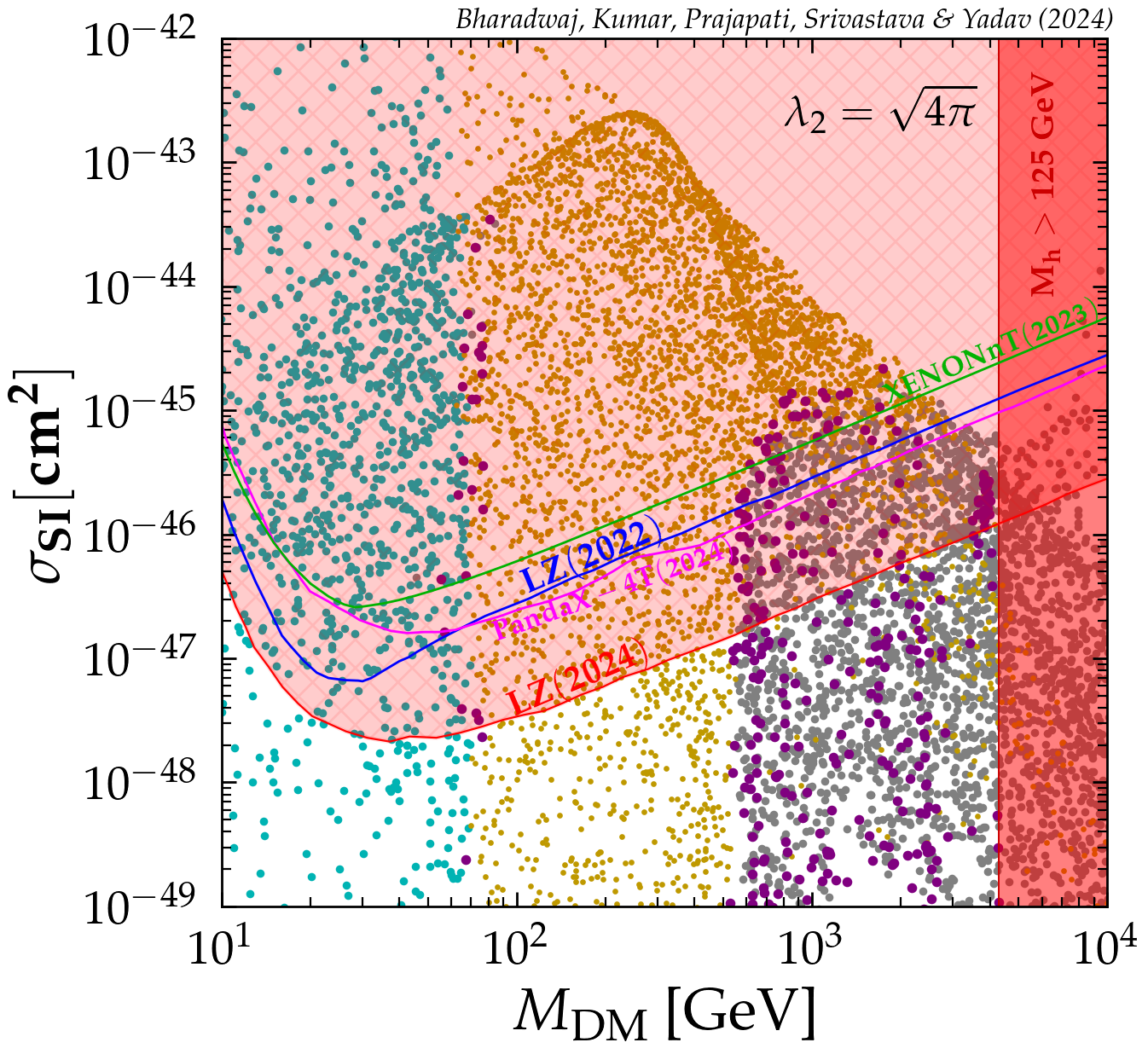}\\
        \caption{Spin-independent nucleon-DM scattering cross-section as a function of DM mass. Higgs mass is computed at two-loop-level with $\lambda_2$ $\approx$ 0 (left) and $\lambda_2$ = $\sqrt{4\pi}$ (right).}
        \label{fig:Inertpb_With_Loop}
\end{figure}

In Fig.~\ref{fig:Inertpb_With_Loop} we plot the spin-independent WIMP-nucleon scattering cross-section as a function of the DM mass, where Higgs mass is computed at the two-loop-level, with two scenarios depicted: $\lambda_2$ $\approx$ 0 in the left panel, and $\lambda_2$ = $\sqrt{4\pi}$ in the right panel.
The constraints from various direct detection experiments
are represented by different colored lines. The cyan-colored points are excluded by various collider constraints, including those from LEP and LHC experiments, as detailed in \cite{Belyaev:2016lok,Cao:2007rm,ATLAS:2023tkt}. 

The direct detection cross-section depends on the magnitude of the $\lambda_{345}$ coupling. If specific values for the $\lambda_{3}$, $\lambda_{4}$ and $\lambda_{5}$ couplings are chosen such that $\lambda_{345}$ is reduced in magnitude, the direct detection cross-section will also decrease accordingly. To maintain consistency with the observed relic density, DM mass can be adjusted as needed. The plot highlights two distinct mass ranges that remain consistent with all experimental bounds including latest direct detection bounds: one a small range near  $M_{h}/2$  and another extending from approximately 500 GeV to 4.4 TeV. In this case, co-annihilation channels are very important while determining the final parameter space. Recall that in the case of real and complex singlet scalar DM, the parameter space is nearly entirely ruled out by the latest LZ (2024) result. In contrast, for the inert doublet model, the presence of co-annihilation channels allows for a finite region of parameter space that remains viable and satisfies all experimental constraints. This feature is also discussed in the Dirac scotogenic model \cite{CentellesChulia:2024iom}, where co-annihilation channels play a crucial role in determining the parameter space.

Thus, a finite region of parameter space remains viable under all experimental constraints for this scenario. However, the DM mass is still constrained to a maximum of approximately 4.4 TeV  beyond which the loop corrections to the Higgs mass simply become too large. Notably, these Higgs mass constraints impose stricter bounds compared to naive perturbativity limits, reducing the maximum allowed parameter space from around 32 TeV to approximately 4.4 TeV.
\FloatBarrier


\section{Conclusions} \label{sec:conc}

In this work, we have analyzed the constraints on DM mass 
coming from DM-induced loop corrections to the Higgs mass.
Our results are based on two key observations: First, in many DM models, the Higgs-mediated channel significantly contributes to satisfying the DM relic density. Second, the same interaction provides loop corrections to the mass of the Higgs boson. We show that an increase in the DM mass necessitates a stronger Higgs-DM coupling to satisfy the relic density, thereby imposing a lower bound on this coupling. However, higher DM mass and stronger Higgs-DM coupling lead to significant DM-induced loop corrections to the Higgs mass. Beyond a certain value of DM mass, these loop corrections become so large that the loop corrected  Higgs mass cannot be maintained at its experimentally observed value.  This imposes a stringent upper bound on the DM mass, typically around a few TeVs.
We argued that such limits are applicable to all kind of non-SUSY DM i.e. scalar, vector and fermionic DM, as long as the DM-Higgs coupling can be constrained from DM relic density computation. 

We further analyzed the implications of these bounds by  considering three simple and well studied DM models, namely, the singlet scalar DM models–real and complex, as well as the inert doublet DM model.
In these models, the Higgs-DM interaction not only determines the DM relic density but also its experimental direct detection prospects. This is because in these models the WIMP-nucleon cross-section is directly proportional to the Higgs-DM interaction strength. 
We find that for the real and complex singlet cases, after imposing the constraints from  Higgs mass correction, relic abundance and direct detection, the entire DM parameter space, except a narrow region near $M_h/2$, is ruled out by the recent LZ (2024) results. In contrast, in the inert doublet model due to the presence of co-annihilation channels, the parameter space up to 4.4 TeV is still allowed. Our findings provide a foundation for extending this analysis to more complex models, which will be addressed in future work.

Before ending we like to mention that our results are rather conservative, in the sense that, we have assumed that DM is the only heavy, $\mathcal{O}$(TeV), BSM particle coupling with Higgs. In many DM models, there is an extended dark sector with DM being the lightest dark sector particle. In such cases, if the other dark sector particles also couple with Higgs, then they will also contribute to the loop corrections to the Higgs mass. Therefore, the Higgs mass constraints will lead to an even stronger constraint on the allowed DM mass. The obvious  exception will be the case of SUSY models. Another potential exception can arise if the dark sector has both fermions and scalars whose coupling with Higgs and their masses are fine tuned such that their loop corrections cancel each other out. But without SUSY such a scenario will require artificial fine tuning.  Thus, in summary our results obtained here are rather conservative and quite robust. They are applicable to a large variety of non-SUSY DM models.

       
\section{Acknowledgments}
\noindent

We are greatly indebted to Prof. Werner Porod for many helpful discussions and for clarifying comments on the SARAH and SPheno prescriptions as well as the manuscript.
His insights and helpful comments have been instrumental in enhancing our clarity and helping us finish this work. We also thank Prof. Martin K. Hirsch and Prof. Subhendra Mohanty for their helpful comments on our manuscript. PB, RK, and SY would like to acknowledge the funding support by the CSIR SRF-NET fellowship. The work of HP is supported by the Prime Minister Research Fellowship (ID: 0401969). \\

\textit{ RS dedicates this work to his PhD students Anirban, Hemant, Praveen, Ranjeet and Sushant who have never shied away from any challenge thrown their way.  As they near the completion of their doctoral journey, their presence will be sorely missed by me.  I wish them best of luck  for their future endeavors. }
 
       
\appendix  
\section{ WIMP–nucleon Scattering Cross-Section For Simple Models}
\label{appendixA}

Here, we discuss the WIMP--nucleon spin-independent scattering cross-section for the simplest cases of scalar, vector, and fermionic DM candidates.\\

\textbf{SCALAR DM :} The Lagrangian density corresponding to the scalar case is,
\begin{align}
-{\cal L}_{\rm Scalar} & \supset
\mu_S^2\, S^2 + \frac{\lambda_{HS}}{2}\,S^2\,H^\dagger H+\frac{\lambda_S}{2}\, S^4\,.
\label{eq:eftsdm2}
\end{align}
The  WIMP–nucleon spin-independent scattering cross-section for this case is given as,
\begin{eqnarray}
 \sigma_{\rm Scalar} & = & \frac{\lambda_{HS}^2 }{4\pi M^4_h}\,\frac{m^4_n}{(m_n + m_{\rm DM})^2} \, f^2_n \,.
\end{eqnarray}

\textbf{VECTOR DM :} Similarly, the Lagrangian density corresponding to the vector DM case can be written as,
\begin{align}
-{\cal L}_{\rm Vector} & \supset 
\frac{1}{2}\, \mu_V^2 V_\mu V^\mu + \frac{1}{4}\, \lambda_{HV} H^\dagger H {V_\mu V^\mu}
+ \frac{1}{4} \lambda_V (V_\mu V^\mu )^2 
\label{eq:eftVec2}\,.
\end{align}
The scattering cross-section, in this case is,

\begin{eqnarray}
 \sigma_{\rm Vector} & = & \frac{\lambda_{HV}^2}{16\pi M^4_h}\,\frac{m^4_n}{(m_n + m_{\rm DM})^2} \, f^2_n \ .
\end{eqnarray}

\textbf{FERMIONIC DM :} Finally, the effective fermionic DM case could be written as,
\begin{align}
-{\cal L}_{\rm Fermion} & \supset \mu_\psi \overline{\psi} \psi 
+ \,\frac{\lambda_{H\psi}}{\Lambda} H^\dagger H ~\overline{\psi} \psi
\label{eq:eftFer2}\ .
\end{align}
The scattering cross-section in this case is,
\begin{eqnarray}
 \sigma_{\rm Fermion} & = & \frac{\lambda_{H\psi}^2}{4\pi M^4_h\,\Lambda^2}\,\frac{m^4_n\, m_{DM}^2}{(m_n + m_{\rm DM})^2} \, f^2_n \ .
 \label{dirdf2}
\end{eqnarray}

 Here, $m_n$ is the nucleon mass, $m_{\rm DM}$ is the DM mass, $M_h$ is the Higgs mass, and  $f_n$ is the Higgs-nucleon effective coupling.



\bibliographystyle{utphys}
\bibliography{references} 
\end{document}